\documentclass[a4paper,fleqn]{cas-dc}

\usepackage{multirow}

\def\tsc#1{\csdef{#1}{\textsc{\lowercase{#1}}\xspace}}
\tsc{WGM}
\tsc{QE}
\tsc{EP}
\tsc{PMS}
\tsc{BEC}
\tsc{DE}

\usepackage{lscape}
\usepackage[dvipsnames]{xcolor}
\usepackage{algorithm}
\usepackage{algorithmic}
\usepackage{caption}
\usepackage{pdflscape}
\usepackage[inline]{enumitem}
\usepackage{changepage}
\usepackage{tabularx}
\usepackage{array}
\usepackage[square,sort,comma,numbers]{natbib}
\setcitestyle{square}
\begin{document}
\let\WriteBookmarks\relax
\def\floatpagepagefraction{1}
\def\textpagefraction{.001}
\newcommand{\subsubsubsection}[1]{\paragraph{#1}\mbox{}\\}
\setcounter{secnumdepth}{4}
\setcounter{tocdepth}{4}
\newcommand{\subsubsubsubsection}[1]{\subparagraph{#1}\mbox{}\\}
\setcounter{secnumdepth}{5}
\setcounter{tocdepth}{5}
% Short title
\shorttitle{Modern Computing: Vision and Challenges}

% Short author
\shortauthors{S S Gill et al.}

% Main title of the paper
\title [mode = title]{Modern Computing: Vision and Challenges}

\author[1]{Sukhpal Singh Gill}[orcid=0000-0002-3913-0369]
\affiliation[1]{organization={School of Electronic Engineering and Computer Science, Queen Mary University of London, London, UK}}
\ead{s.s.gill@qmul.ac.uk}

\author[2]{Huaming Wu}
\affiliation[2]{organization={Center for Applied Mathematics, Tianjin University, Tianjin, China}}
\ead{whming@tju.edu.cn}

\author[3]{Panos Patros}
\affiliation[3]{organization={Raygun Performance Monitoring, Wellington, New Zealand}}
\ead{panos@raygun.com}

\author[4]{Carlo Ottaviani}
\affiliation[4]{organization={Department of Computer Science and York Centre for Quantum Technologies, University of York, York, UK}}
\ead{carlo.ottaviani@york.ac.uk }

\author[5]{Priyansh Arora}
\affiliation[5]{organization={Microsoft, Schiphol, Netherlands}}
\ead{priyansh.arora@microsoft.com }

\author[6]{Victor Casamayor Pujol}
\affiliation[6]{organization={ Distributed Systems Group, Vienna University of Technology, Vienna, Austria}}
\ead{v.casamayor@dsg.tuwien.ac.at }

\author[7]{David Haunschild}
\affiliation[7]{organization={ Detecon International GmbH, Munich, Germany}}
\ead{david.haunschild@detecon.com }

\author[8]{Ajith Kumar Parlikad}
\affiliation[8]{organization={ Institute for Manufacturing, Department of Engineering, University of Cambridge, Cambridge, UK}}
\ead{ aknp2@cam.ac.uk}

\author[9]{Oktay Cetinkaya}
\affiliation[9]{organization={Oxford e-Research Centre (OeRC), Department of Engineering Science, University of Oxford, Oxford, UK}}
\ead{oktay.cetinkaya@eng.ox.ac.uk }

\author[10]{Hanan Lutfiyya}
\affiliation[10]{organization={Department of Computer Science, University of Western Ontario, London, Canada}}
\ead{hanan@csd.uwo.ca }

\author[11]{Vlado Stankovski}
\affiliation[11]{organization={Faculty of Computer and Information Science, University of Ljubljana, Ljubljana, Slovenia}}
\ead{vlado.stankovski@fri.uni-lj.si }

\author[12]{Ruidong Li}
\affiliation[12]{organization={Institute of Science and Engineering, Kanazawa University, Japan}}
\ead{lrd@se.kanazawa-u.ac.jp}

\author[13]{Yuemin Ding}
\affiliation[13]{organization={Tecnun School of Engineering, University of Navarra, Spain}}
\ead{yueminding@tecnun.es}

\author[14]{Junaid Qadir}
\affiliation[14]{organization={Department of Computer Science and Engineering, College of Engineering, Qatar University, Doha, Qatar}}
\ead{jqadir@qu.edu.qa}

\author[15,16]{Ajith Abraham}
\affiliation[15]{organization={Bennett University, Greater Noida, India}}
\affiliation[16]{organization={Machine Intelligence Research Labs, Auburn, WA, USA}}
\ead{ ajith.abraham@ieee.org}

\author[17]{Soumya K. Ghosh}
\affiliation[17]{organization={ Department of Computer Science and Engineering, Indian Institute of Technology, Kharagpur, India}}
\ead{ skg@cse.iitkgp.ac.in}

\author[18]{Houbing Herbert Song}
\affiliation[18]{organization={ Department of Information Systems University of Maryland, Baltimore County (UMBC), Baltimore, USA}}
\ead{ songh@umbc.edu}

\author[19]{Rizos Sakellariou}
\affiliation[19]{organization={ Department of Computer Science, University of Manchester, Oxford Road, Manchester, UK}}
\ead{rizos@manchester.ac.uk}

\author[20]{Omer Rana}
\affiliation[20]{organization={School of Computer Science and Informatics, Cardiff University, Cardiff, UK}}
\ead{ranaof@cardiff.ac.uk}

\author[21, 22]{Joel J. P. C. Rodrigues}
\affiliation[21]{organization={Department of Computer Science, College of Computer and Information Sciences, King Saud University, Riyadh, Saudi Arabia}}
\affiliation[22]{organization={COPELABS, Lusófona University, Lisbon, Portugal}}
\ead{joeljr@ieee.org}

\author[23]{Salil S. Kanhere}
\affiliation[23]{organization={School of Computer Science and Engineering, The University of New South Wales (UNSW), Sydney, Australia}}
\ead{salil.kanhere@unsw.edu.au}

\author[6]{Schahram Dustdar}
\ead{ dustdar@dsg.tuwien.ac.at}

\author[1]{Steve Uhlig}
\ead{steve.uhlig@qmul.ac.uk}

\author[24]{Kotagiri Ramamohanarao}
\affiliation[24]{organization={Retired Professor, The University of Melbourne, Victoria, Australia}}
\ead{rkotagiri@gmail.com}

\author[25]{Rajkumar Buyya}
\affiliation[25]{organization={Cloud Computing and Distributed Systems (CLOUDS) Laboratory, School of Computing and Information Systems, The University of Melbourne, Australia}}
\ead{rbuyya@unimelb.edu.au}

\cortext[cor1]{Correspondence to: School of Electronic Engineering and Computer Science, Queen Mary University of London, London, E1 4NS, UK.}

\begin{abstract} Over
\textcolor{black}{the past six decades, the computing systems field has experienced significant transformations, profoundly impacting society with transformational developments, such as the Internet and the commodification of computing. Underpinned by technological advancements, computer systems, far from being static, have been continuously evolving and adapting to cover multifaceted societal niches. This has led to new paradigms such as cloud, fog, edge computing, and the Internet of Things (IoT), which offer fresh economic and creative opportunities. Nevertheless, this rapid change poses complex research challenges, especially in maximizing potential and enhancing functionality. As such, to maintain an economical level of performance that meets ever-tighter requirements, one must understand the drivers of new model emergence and expansion, and how contemporary challenges differ from past ones. To that end, this article investigates and assesses the factors influencing the evolution of computing systems, covering established systems and architectures as well as newer developments, such as serverless computing, quantum computing, and on-device AI on edge devices. Trends emerge when one traces technological trajectory, which includes the rapid obsolescence of frameworks due to business and technical constraints, a move towards specialized systems and models, and varying approaches to centralized and decentralized control. This comprehensive review of modern computing systems looks ahead to the future of research in the field, highlighting key challenges and emerging trends, and underscoring their importance in cost-effectively driving technological progress.}

\end{abstract}

% Research highlights
% \begin{highlights}
% \item We explore the evolution of computing paradigms \& technological drivers (1960 onward).
% \item We offer a taxonomy of modern computing based on impact and performance criteria.
% \item We classify computing based on paradigms, technologies, impact areas, and trends.
% \item We identify open challenges and research directions for computing traits.
% \item We introduce a hype cycle for modern computing systems, spotlighting emerging trends.

% \end{highlights}

\maketitle
\section{Introduction}
\label{introduction}
The Internet, the expansive computational backbone of interactive machines, is largely responsible for the 21st century's social, financial, and technological growth~\cite{buyya2018manifesto}. The growing reliance on the computing resources it encapsulates has pushed the complexity and scope of such platforms, leading to the development of innovative computing systems. These systems have genuinely improved the capabilities and expectations of computing equipment driven by rapid technical and user-driven evolution~\cite{lindsay2021evolution}. For instance, vintage mainframes combined centralized data processing and storage with transmission interfaces for user input. Due to advancements in clusters and packet-switching technologies, microchip gadgets, and graphical user interfaces, technology originally shifted from big, centrally-run mainframe computers to Personal Computers (PCs). The globalization of network standards made it possible for interconnected networks worldwide to communicate and share data~\cite{yamashita2020history}. Businesses slowly combined sensor and actuator goals with built-in network connectivity by creating architectures and standards that submit tasks to remote pools of computing resources, such as memory, storage, and data processing~\cite{gill2022ai}. As a result, newer models like the Internet of Things (IoT) and edge computing are now beginning to expand the reach of technology outside the confines of traditional network nodes~\cite{gubbi2013internet}.

Over the past six decades, computing models have fundamentally shifted to address the problems posed by the ever-evolving nature of our civilization and its associated computer system architectures~\cite{muralidhar2022energy}. The evolution of computing from mainframes to workstations to the cloud to autonomous and decentralized architectures, such as edge computing and IoT technologies, however, maintains identical core parts and traits that characterize their function~\cite{chakraborty2023journey}. Research in computing underpins all of them! Advancements in areas like security, computer hardware acceleration, edge computing, and energy efficiency typically serve as catalysts for innovation and entrepreneurship that span across various business domains~\cite{beloglazov2012energy}. While computing systems and other forms of system integration create new problems/opportunities, software frameworks have been developed to address them. Thus, middleware, network protocols, and safe segregation techniques must be continually developed and refined to support novel computing systems---and their innovative use cases.

\subsection{Motivation}
By tracking the effect of computing systems on the community, this comprehensive study seeks to (a) establish the essential features and components of modern computing systems, (b) thoroughly assess the development of innovations and behavioral patterns that inspired the invention of these paradigms, and (c) recognize significant developments throughout the models, such as the integration of system design, the shifting between centralization and decentralization, and lags in model conceptualization and development.

This investigation suggests that next-generation computing systems will facilitate the decentralization of computational services. This will be achieved via the composition of decentralized calculation tools with workload-specific targets for performance to create dramatically more complex structures. These will satisfy holistic operational demands, such as improved capacity and power accessibility. 

\subsection{\textcolor{black}{Related Surveys and Our Contributions}}

\textcolor{black}{Computing being a rapidly growing topic, the time is right for a novel, forward-thinking study to summarize, improve, and integrate the existing and newly-generated information, and to explore possible trends and future viewpoints. Previously, Pujol et al.~\cite{casamayor2023fundamental} provided a survey on distributed computing continuum systems that focused on business models. Further back in 2018, Buyya et al.~\cite{buyya2018manifesto} presented a manifesto on fundamental issues, developments, and impacts in cloud computing research.  Meanwhile, Gill et al.~\cite{gill2022ai} offered a visionary survey of advances in computing paradigms for fog, edge, and serverless computing. Further, Shalf~\cite{shalf2020future} summarized the 2020 state of the art of technological roadmaps and their implications for the future of systems, including what a post-exascale system would entail. Finally, in 2021, Angel et al.~\cite{angel2021recent} reviewed leading computational frameworks for cloud and edge computing, and showcased breakthroughs that had been brought about via the merging of Machine Learning (ML) with these models.}

\textcolor{black}{In order to evaluate and identify the most pressing research issues of modern computing, we have developed the very first taxonomy of its type. We performed a gap analysis of the current surveys using several criteria, as shown in Table \ref{table:comparison}, which underpinned the design of our work. Hence, our study uniquely contributes by (a) exploring the history of computing paradigm shifts with a focus on technology drivers, (b) providing a thorough taxonomy of computing systems, (c) introducing the hype cycle for modern computing systems with a focus on new trends, and (d) discussing the effects and cost-effective performance requirements of modern computing.}

\textbf{\begin{table*}[ht!]
\caption{\textcolor{black}{Comparison of this work with existing studies}}
 \centering
\small \resizebox{1\textwidth}{!}{
\begin{tabular}{|cl|l|l|l|l|l|l|}
\hline
\multicolumn{2}{|l|}{Work} & \cite{casamayor2023fundamental} & \cite{buyya2018manifesto}  & \cite{gill2022ai} & \cite{shalf2020future} & \cite{ angel2021recent} & Our Work\\ \hline
\multicolumn{2}{|l|}{Year}  & 2023 & 2018 & 2022 & 2020 & 2021 &   2024 \\ \hline
\multicolumn{2}{|l|}{A Taxonomy of Modern Computing} &  &  &  &     &  & \checkmark \\ \hline
\multicolumn{2}{|l|}{Evolution of Computing Paradigms (1960 to 2023)}  &  & &  &  &  & \checkmark \\ \hline
\multicolumn{1}{|c|}{\multirow{4}{*}{\begin{tabular}[c]{@{}c@{}} Classification of Computing\end{tabular}}} & \begin{tabular}[c]{@{}l@{}}Standalone vs. \\ Networked Computing\end{tabular}&  &  &   &   &  & \checkmark \\ \cline{2-8} 
\multicolumn{1}{|c|}{} & \begin{tabular}[c]{@{}l@{}}General Purpose vs. \\ Specialized Computing\end{tabular} &   & &  &  &  & \checkmark\\ \cline{2-8} 
\multicolumn{1}{|c|}{} & \begin{tabular}[c]{@{}l@{}}Centralized vs. \\ Decentralized Computing\end{tabular} &  &  &  &  &  & \checkmark \\ \cline{2-8} 
\multicolumn{1}{|c|}{} & \begin{tabular}[c]{@{}l@{}}Computing Trends and \\ Emerging Technologies\end{tabular}    & \checkmark & \checkmark & \checkmark & \checkmark & \checkmark &   \checkmark \\ \cline{2-8} 
\multicolumn{1}{|c|}{} & \begin{tabular}[c]{@{}l@{}}Computational Methodologies: \\ Parallel vs. Sequential Computing\end{tabular}  &  &  &  &  &  &   \checkmark \\ \cline{2-8}   \hline
\multicolumn{1}{|c|}{\multirow{3}{*}{Traits of Computing}} & Focus/ Paradigms &  &  \checkmark &  &  & &  \checkmark  \\  \cline{2-8}  
\multicolumn{1}{|c|}{} & Technologies/ Impact Areas &  &  &  &  &  &   \checkmark \\ \cline{2-8}
\multicolumn{1}{|c|}{} & Trends/ Observations &  &  &  &  &  &  \checkmark \\ \cline{2-7} \hline
\multicolumn{1}{|c|}{\multirow{5}{*}{Impact and Performance Criteria}} & Performance Metrics &   & &  &  &  & \checkmark \\ \cline{2-8}
\multicolumn{1}{|c|}{} & Efficiency Metrics &  &  &  &  &  &  \checkmark \\ \cline{2-8}
\multicolumn{1}{|c|}{} & \begin{tabular}[c]{@{}l@{}}Social Impact \end{tabular}  &  &  &  &  &  & \checkmark \\ \multicolumn{1}{|c|}{} & Security and Compliance  &  &  &  &  &  &    \checkmark \\ \cline{2-8} 
\multicolumn{1}{|c|}{} & \begin{tabular}[c]{@{}l@{}}Economic and Management \end{tabular}   &  &  &  &  &  & \checkmark  \\\hline
\multicolumn{2}{|l|}{Open Challenges and Future Directions}  & \checkmark & \checkmark & \checkmark & \checkmark & \checkmark &   \checkmark \\ \hline
%\multicolumn{2}{|l|}{Further Reading}  &  &  &  &  &  &   \checkmark  \\ \hline
 \multicolumn{2}{|l|}{Emerging Trends in Modern Computing: Hype Cycle } &  & & &  &  &   \checkmark  \\ \hline
\end{tabular}
}
\label{table:comparison} 
\end{table*}}

\textcolor{black}{The \textbf{key contributions} of this article are summarized as follows:}

\begin{itemize}
    \item  \textcolor{black}{It offers a concise overview of the transition from early to modern computing.}
    \item  \textcolor{black}{The study explores the evolution of computing paradigms, focusing on technological drivers (1960–2023).}
    \item \textcolor{black}{Following a novel methodology, the article produces a taxonomy of modern computing based on traits of computing such as 1) focus or paradigms; 2) technologies or impact areas; and 3) trends or observations.  }
\item \textcolor{black}{It presents a comprehensive classification of computing: 1) Standalone vs. Networked Computing; 2) General Purpose vs. Specialized Computing, 3) Centralized vs. Decentralized Computing, 4) Computing Trends and Emerging Technologies; and 5) Computational Methodologies: Parallel vs. Sequential Computing.}
    \item \textcolor{black}{The study identifies the impact and performance criteria of modern computing in terms of performance metrics, efficiency metrics, social impact, security and compliance, and economics and management.}
       \item \textcolor{black}{It provides an in-depth summary of computing traits and resources for further research.}
      \item  \textcolor{black}{The article identifies open challenges and research directions for the traits of computing.}
\item \textcolor{black}{Finally, it introduces the hype cycle for modern computing systems, spotlighting emerging trends.}
\end{itemize}

\subsection{Article Organization}
The article is organized as follows: Section~\ref{early} offers a concise overview of the transition from early to modern computing. Section~\ref{evolution} explores the evolution of computing paradigms, focusing on technological drivers. Section~\ref{classification} presents a classification of computing systems, and Section~\ref{Impact} examines the impact and performance criteria in modern computing. The article concludes in Section~\ref{Conclusion}, summarizing computing-related technologies and trends through a hype cycle in Section~\ref{Emerging}. The list of acronyms used in this study is given in Appendix~A.

\section{Early Computing to Modern Computing: A Vision} 
\label{early}
\begin{landscape}
\begin{figure}[t!]
    \centering
    \includegraphics[scale=0.40,angle=0]{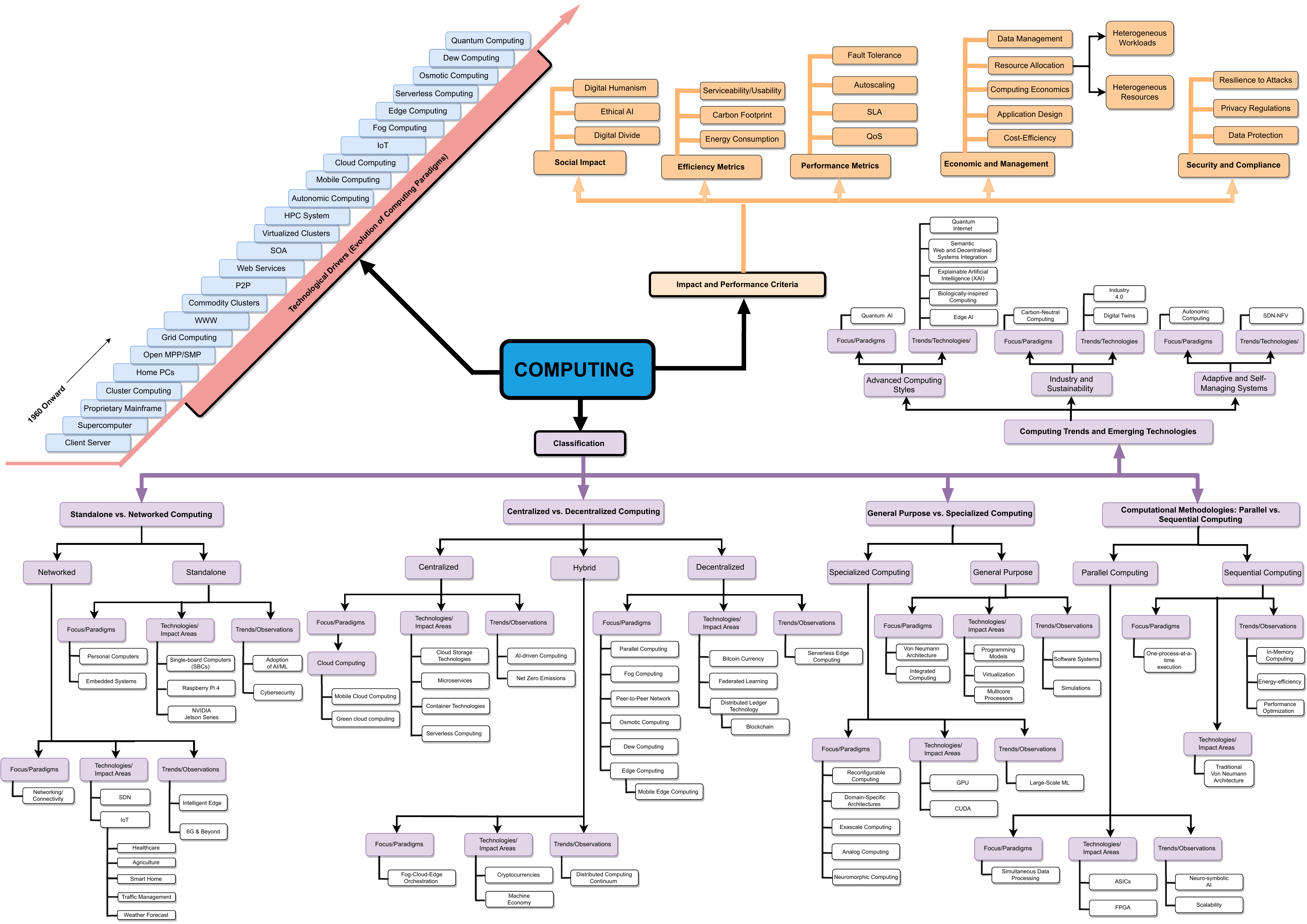}
    \caption{Modern Computing: A Taxonomy}
    \label{fig:tax}
\end{figure}
\end{landscape}

Over the last six decades, advancements in computing systems have optimized the efficiency of the available hardware~\cite{rimal2009taxonomy}. Over this time period, novel computing models and innovations have been developed and replaced the previous state-of-the-art, all of which incrementally contribute to the current technology status~\cite{lindsay2021evolution}. Fig.~\ref{fig:tax} shows the transition from early computing to contemporary computing. Originally, a single system could only carry out a single task; hence, a user needed various systems working in tandem to achieve their desired tasks. However, to safely share information between computers---in order to overcome the problem of executing only one task at a time---a reliable communication mechanism is essential~\cite{gill2019transformative}. To that end, our investigation unfolds across three key sections: Section~\ref{evolution} delves into the evolution of computing paradigms, emphasizing technological drivers. Section~\ref{classification} offers a comprehensive classification of computing systems. The discussion in Section~\ref{Impact} revolves around the impact and performance criteria of modern computing. Section~\ref{Emerging} introduces the hype cycle for modern computing systems, spotlighting emerging trends.

%\begin{itemize}

\section{Evolution of Computing Paradigms: Technological Drivers}
\label{evolution}

Figure~\ref{fig:tax} illustrates the progression of computing technology starting from the year 1960.
    
%\item 
\subsection{Client Server}
In the year 1960, a centralized platform (a.k.a, distribution integration) was developed to share workloads (a.k.a., jobs) between the resource providers (i.e., server instances) and service consumers (i.e., customers)~\cite{rimal2009taxonomy}. Supporting it, a networking system was utilized for communications between client devices and servers, and servers exchange resources for customers to perform their tasks using a load balancing mechanism~\cite{flynn1966very}. Illustrative examples of the client-server model's application include the Email and the World Wide Web (WWW). However, users in this configuration were unable to freely interact with one another.

\subsection{Supercomputer}
A supercomputer is a powerful computer with extraordinary processing capability, such that it can handle complex calculations in several areas of science, including climate study, quantum physics, and molecular simulation~\cite{kozyrakis1998new}. Energy utilization and heat control in supercomputers endured as a key research problem throughout their growth in the 1960s~\cite{casavant1988taxonomy}. Supercomputers, such as Multivac, HAL-9000, and Machine Stops, have been instrumental in underpinning/enabling dramatic technological advancements~\cite{flynn1966very}.

\subsection{Proprietary Mainframe}
To handle massive amounts of data (including dealing with transactions, customer data analysis, and censuses), a high-speed machine with large computing power is required~\cite{yu2005taxonomy}. Virtualization on mainframes allows for increased efficiency, protection, and dependability. In the year 2017, IBM announced the newest version of its mainframe, the IBM z14~\cite{gill2019transformative}. Being built to support massive economic activity and despite their high price tag, mainframe computers deliver outstanding efficiency~\cite{flynn1966very}.

\subsection{Cluster Computing} 
Cluster computing is a method of increasing the efficiency of a computing system by utilizing several nodes to complete a single operation~\cite{owens2008gpu}. In order to coordinate various computing nodes, this type of technology requires a rapid Local Area Network (LAN) for exchanging information among them~\cite{compton2002reconfigurable}.

\subsection{Home PCs}
The early days of the Internet coincided with the flourishing of Personal Computers (PC) kept at one's home~\cite{yamashita2020history}. The Internet was evolving into a foundational network, connecting local networks to the larger Internet using self-adaptive network protocols, such as Transmission Control Protocol/Internet Protocol (TCP/IP)---in contrast to the original Network Control Protocol (NCP)-based Advanced Research Projects Agency Network (ARPANET) mechanisms~\cite{lindsay2021evolution}. As a result, there was a sharp increase in the number of hosts on the Internet, which quickly overwhelmed centralized naming technologies like \textit{HOSTS.TXT}. In the year 1985, the earliest publicly available version of a Domain Name System (DNS) was released for the Unix BIND system~\cite{wright2010cybersquatting}. This system translates hostnames into IP addresses. Pioneer Windows,
Icons, Menus, and Pointers (WIMP)-based Graphical User Interfaces (GUIs) on computers, such as the Xerox Star and the Apple LISA, proved that customers could successfully use machines in their homes for tasks like playing video games and surfing the Internet~\cite{jansen1998graphical}.

\subsection{Open MPP/SMP}
Massive Parallel Processing (MPP) and Symmetric Multi-Processing (SMP) systems are the two most common forms of parallel computing platforms~\cite{casavant1988taxonomy}. In an SMP setup, multiple processors run the same Operating System (OS) concurrently while sharing the rest of the hardware's capacity (e.g., disc space and RAM). Naturally, resource pooling influences the computational speed of completing a given assignment. In an MPP scenario, the file system can be shared, while no other resources are pooled for use during task processing~\cite{flynn1966very}. Incorporating more machines~$N$ and their associated storage and RAM space, increases the ability to scale according to the Universal Scalability Law (an extension of Amdahl's Law), assuming $\kappa$ the proportion of work that can be parallelized and $\sigma$ the interprocess communication penalty:
\begin{equation}
    \text{Capacity} = \frac{N}{1 + \sigma \cdot (N - 1) + \kappa \cdot N \cdot (N - 1)}. 
\end{equation}

\subsection{Grid Computing} 
This technology enables a group to work together towards the same objective by executing non-interactive, and largely IO-intensive tasks~\cite{compton2002reconfigurable}. Each application running on only one grid is a top priority~\cite{rimal2009taxonomy}. In addition to allocating and managing resources, grid computing also offers a reliable architecture, as well as tracking and exploration support.

\subsection{WWW} 
The primary web browsers, websites, and web servers all came into existence in the later stages of the 1980s and early 1990s, underpinned by the development of Hyper Text Transport Protocol (HTTP) and Hyper Text Markup Language (HTML)~\cite{lindsay2021evolution}. The platform for the interconnected system of networks that makes up the WWW was made possible by the standardizing technology of TCP/IP network protocols. This allowed for a dramatic increase in the total number of servers linked to the Web and introduced Information Technology (IT) to the general public. Software applications were thus able to communicate with one another beyond address spaces and networking, e.g., via novel technologies like Remote Procedure Calls (RPCs)~\cite{tay1990survey}.

\subsection{Commodity Clusters} 
Commodity cluster computing employs several computers simultaneously, which can inexpensively execute user tasks~\cite{compton2002reconfigurable}. In an effort to standardize their processes, several companies use open standards while building commodity computers~\cite{flynn1966very}. This allowed immediate computing business needs to be met using ready-made processors.

\subsection{Peer to Peer (P2P)} 
P2P is a distributed framework to share workloads or jobs amongst multiple peers; alternatively, computers and peers may interact with one another openly at the application layer~\cite{suryono2019peer}. With no mediator in the center, users of a peer-to-peer system can share resources like memory, CPU speed, and storage space. Peer-to-peer communication utilizes the TCP/IP protocol suite~\cite{schollmeier2003protocol}. Interactive media, sharing file infrastructure, and content distribution are some of the most common use cases for P2P technology.

\subsection{Web Services} 
The technology supporting web services enables the exchange of data between various Internet-connected devices in machine-understandable data formats, such as JavaScript Object Notation (JSON) and Extensible Markup Language (XML), over the WWW~\cite{alonso2004web}. Commonly, web-based services operate as a connection between end users and database servers.

\subsection{Service-Oriented Architecture (SOA)}  
The SOA paradigm enables software elements to be reused and made compatible through advertised service designs/APIs~\cite{perrey2003service}. It is normally easier to include services in new apps: the apps can be architected to adhere to standardized protocols and leverage consistent design patterns. This frees the software engineer from the burden of recreating or duplicating current features or figuring out how to link to and interoperate with current systems---e.g., via using Software Development Kits (SDKs) that implement common functionalities, such as networking, retries, marshalling of data and error handling~\cite{maffione2016software}. Each SOA API exposes the logic and data necessary to carry out a single, self-contained business operation (such as vetting the creditworthiness of a client, determining the loan's due date, or handling an insurance application)~\cite{resende2007handling}. The loose integration provided by the service's design allows for the service to be invoked with limited knowledge of the underlying service implementation.

\subsection{Virtualized Clusters}
Virtualization enables a guest computer system to be implemented on top of a host computer system, which abstracts away the problem of physically supporting and maintaining multiple types/architectures of physical machines~\cite{compton2002reconfigurable}. With a virtualized cluster, several Virtual Machines (VMs) may pool their resources to complete a single job. VM hypervisors, which execute the guest system on the host system, allow software-based virtualization to run either on top of an OS or directly (bare-metal) on hardware~\cite{flynn1966very}. Costs and complexity are reduced, and a greater number of tasks may be completed with identical hardware by adopting a VM-based system.

\subsection{High Performance Computing (HPC) System}
HPC is the computing method of choice when dealing with computationally intensive issues, which tend to arise in the domains of commerce, technology, and research~\cite{compton2002reconfigurable, flynn1966very}. A scheduler in an HPC system manages accessibility to the various computing resources available for use in solving various issues~\cite{mergen2006virtualization}. HPC systems utilize a pooled set of resources, allowing them to perform workloads or tasks via the allocation of concurrent resources and online utilization of various resources.

\subsection{Autonomic Computing} 
One of the first global initiatives to build computer systems with minimum human involvement to achieve preset goals was IBM's autonomic computing program in 2006~\cite{kephart2003vision}. It was mostly based on research on nerves, thinking, and coordination. Autonomic computing research examines how software-intensive systems may make choices and behave autonomously to achieve user-specified goals~\cite{gill2022ai}. Control for closed- and open-loop systems has shaped autonomic computing~\cite{singh2017star}. Complex systems can have several separate control networks.

\subsection{Mobile Computing} 
The term ``mobile computing'' is used to describe a wide range of IT components that give consumers mobility in their usage of computation, information, and associated equipment and capabilities~\cite{othman2013survey}. An especially popular definition of ``mobile'' is accessing information while moving, when an individual is not confined to a fixed place. Accessibility at a fixed spot may also be thought of as mobile, especially if it is provided by hardware that consumers can move as needed but that remains in one place while functioning~\cite{alahmad2021mobile}. Mobile computing devices are becoming essential across industries, boosting efficiency and creativity in fields such as healthcare, retail, manufacturing, and the arts.

\subsection{Cloud Computing} 
Software as a service (SaaS), platform as a service (PaaS), and infrastructure as a service (IaaS) are all examples of Internet-accessible web services~\cite{buyya2018manifesto}. Google Mail is an excellent instance of a SaaS product since it provides a wide range of useful features without the burden of installation and ongoing upkeep costs. PaaS providers like Microsoft provide a scalable environment where users can install their applications~\cite{anwar2021recommender}. Amazon is a prime instance of an IaaS provider since it provides users with access to servers, networks, storage, and other hardware components necessary to run applications and other workloads efficiently and effectively. Using distant facilities for performing user operations (processing, administration, and storage of data) over the Internet is known as ``cloud computing,'' abbreviated as ``XaaS,'' where X = ``I,'' ``P,'' ``S,'' etc. Cloud computing enables the pooling of resources to reduce execution costs and enhance service accessibility~\cite{durao2014systematic}. There are four major types of cloud computing systems: public, private, hybrid, and communal. Dependability, safety, and cost-effectiveness are just a few examples of Quality of service (QoS) characteristics that should be considered while developing a successful cloud service.

\subsection{IoT} 
Controllers, gadgets, and detection devices are all examples of IoT devices that can communicate with one another over the WWW~\cite{gubbi2013internet}. IoT has many potential uses in many different areas, including farming, medical treatment, climate prediction, logistics, home automation, and industrial automation \cite{gill2019router}.

\subsection{Fog Computing} 
This cutting-edge design makes extensive use of mobile devices, also known as fog nodes, which are utilized for data storage and processing, and rely on the web for inter-node connectivity~\cite{iftikhar2022ai}. The data plane and the control plane are the two main components of fog computing~\cite{gill2019fog}. Although the control layer is a gateway component and determines the network's layout, the data plane offers capabilities at the network's edge to decrease delay and boost QoS~\cite{singh2021fog}. Fog computing supports IoT gadgets such as smartphones, detectors, and health monitors.

\subsection{Edge Computing} 
Edge computing is a method that delegates processing to dispersed edge devices for data processing and information exchange~\cite{shi2016edge}. In addition, edge computing enhances QoS, decreases delay, and lowers transmitting expenses by computing huge volumes of data on gadgets at the edge rather than in the public cloud~\cite{10335918}. However, edge computing relies on a constantly available web connection to perform certain tasks in a timely manner, so it's best used for applications that can execute autonomously without centralized control for prolonged periods of time~\cite{khan2019edge}.

\subsection{Serverless Computing} 
The serverless computing paradigm eliminates the need to manage servers and other infrastructure components~\cite{jonas2019cloud} centrally. Since serverless computing eliminates the need for software engineers to manage servers, it is expected to grow much faster. With serverless computing, hosting companies may easily handle infrastructure management and automatic provisioning~\cite{hassan2021survey}. Because of this, less effort and resources are needed to oversee the infrastructure. 

\subsection{Osmotic Computing} 
Osmotic computing is a growing idea that merges IoT, cloud, fog, and edge technology for the constantly changing administration of IT services. The dramatic increase in the size of resources in the network's periphery is the primary force behind this trend. By defining, creating, and implementing a computing model, this paradigm focuses on methods to improve edge and cloud-based IoT services~\cite{buzachis2021modeling}. To manage resources and resolve data difficulties in IoT and data science, osmotic computing applies the fundamental concepts of the osmosis phenomenon in chemistry~\cite{neha2022systematic}. The primary objective of this computing model is to distribute workloads and efficiently use available resources among servers without degrading service delivery or efficiency. 

\subsection{Dew Computing} 
Dew computing is ``a software-hardware organization model for computers situated in the cloud computing environment,'' where a local machine complements and operates independently of cloud services~\cite{ray2017introduction}. Dew computing may bridge the gap between cloud and on-premises computing. Data and services stored in the cloud are accessible regardless of an Internet connection. The need for constant Internet access is the primary restriction on cloud and fog computing. Complementing fog and edge computing with considerable Internet reliance, an extra layer, including dew computing, is necessary to keep apps and services alive and functioning. Even if dew computing is not conducted entirely online, it nevertheless uses cloud computing and depends on collaboration for data and operations, for example, One Drive~\cite{gushev2020dew}. 

\subsection{Quantum Computing} 
Quantum computing is a radically different way to analyze knowledge and data. Several possibilities can be taken advantage of when processing information stored in the quantum states of quantum machines that are unavailable when analyzing information in a conventional fashion~\cite{qu2020blockchained}. The phenomena of quantum entanglement and superposition are two such examples. Because of quantum entanglement, it is difficult to offer a comprehensive description from the understanding of merely the component states, which is a defining characteristic of quantum systems. One definition of the term ``superposition'' is the potential of merging quantum states to create a new valid quantum state~\cite{kovachy2015quantum}. The primary purpose driving the effort to construct a quantum computer was the modelling of quantum systems; however, it was not until the identification of quantum algorithms capable of achieving realistic objectives that the enthusiasm for constructing such devices began to garner increasing scrutiny~\cite{gill2022quantum}.   
   
%\end{itemize}

\section{Classification of Computing: Paradigms, Technologies and Trends} 
\label{classification}
In this section, we discuss the different types of computing and classify them into different broad categories as shown in Fig.~\ref{fig:tax}. Table~\ref{tab:table1} briefly describes traits of computing that are used in this classification such as 1) focus or paradigms; 2) technologies or impact areas; and 3) trends or observations.

\begin{table*}[!t]
\caption{Summary of Computing Traits}
\label{tab:table1} %shows the 
\centering
\begin{tabular}{p{2.5cm}|p{13.5cm}}
\bottomrule
\textbf{Trait } & \textbf{Description} \\
\hline
Focus/ Paradigms & We discuss well-established computing paradigms, from client-server to quantum computing, which have been explored in the last decade. \\
\hline
Technologies/ Impact Areas &  We cover key research that has grown over time by utilizing these well-established computing paradigms and how this has led to many breakthroughs in the underlying technology.  \\
\hline
Trends/ Observations &  The new trends, such as large-scale machine learning, digital twins, edge AI, bitcoin currency, 6G \& Beyond and quantum Internet and biologically-inspired computing, for the next generation of computing, have come to light due to these advances in computing paradigms and technology.\\
\hline
\end{tabular}
\end{table*}

\subsection{Standalone vs. Networked Computing }
Standalone computing occurs when a computer is not connected to another computer in any way, whether through wired or WiFi connections~\cite{gulliver2004pervasive}. Multiple computers linked together form a network, a model that falls under networked computing.
\subsubsection{Standalone Computing } 
In this section, we discuss the main focus or paradigms, technologies or impact areas, and various trends or observations within standalone computing.

\subsubsubsection{Focus/Paradigms} 
The following are the main focus or paradigms for standalone computing:

\begin{enumerate*}[label =\color{black} \it \arabic*),itemjoin=\\\hspace*{\parindent}]
\item \textit{PCs}: Individuals use PCs, which leverage microprocessors designed for personal use. Before the PC, businesses had to operate computers by connecting several users' terminals to a separate, massive mainframe system~\cite{yamashita2020history}. By the end of the 1980s, technical developments had enabled the construction of a compact computer that a person could purchase and use as a word processor or for various computing objectives~\cite{lindsay2021evolution}.
\item \textit{Embedded Systems}: A computer (often a microcontroller or microprocessor) is built into (i.e., embedded in) the design of a device~\cite{ravi2004security}. Most of the time, an individual does not even realize they are using a computer because there might not be any obvious hints of applications, data, or software~\cite{de2019literature}. The software that operates a microwave oven or an engine control unit of a contemporary vehicle are two instances of items with undetectable integrated systems.
\end{enumerate*}

\subsubsubsection{Technologies/Impact Areas} 
The key technologies and affected domains for standalone computing include:

\begin{enumerate*}[label =\color{black} \it \arabic*),itemjoin=\\\hspace*{\parindent}]
\item \textit{Single-board Computers (SBCs)}: In an SBC, the CPU, I/O, memory, and various other components are all housed on one integrated circuit board; the quantity of memory is fixed; and there are no slots to be expanded for additional hardware~\cite{basford2020performance}.

\item \textit{Raspberry Pi 4}: The Raspberry Pi is a family of tiny SBCs that have been developed to allow programming and computing capabilities to be available to all. The Raspberry Pi Model B became the inaugural board produced by the foundation behind the Raspberry Pi~\cite{basford2020performance}. Due to its immense popularity, other variants have subsequently been developed, each with its own set of advantages. These include the Raspberry Pi computation component, which has been optimized for use in embedded systems~\cite{pajankar2017raspberry}.

\item \textit{NVIDIA Jetson Series}: This is a line of Graphics Processing Units (GPUs) that includes the initial processors built with the explicit purpose of powering self-driving robots~\cite{hwu2017self}. With up to 32 Tera Operations Per Second (TOPS) of Artificial Intelligence (AI) efficiency, these GPUs efficiently handle optical measurements, sensor fusion, positioning, visualization, obstacle detection, and path-planning, all of which are essential for the development of robotics~\cite{basford2020performance}. The Jetson Xavier series focuses on creating specialized robots and edge robots, with several distinct manufacturing components.
\end{enumerate*}

\subsubsubsection{Trends/Observations} 
The main trends and observations regarding standalone computing are:

\begin{enumerate*}[label =\color{black} \it \arabic*),itemjoin=\\\hspace*{\parindent}]
\item \textit{Adoption of AI/ML}: NVIDIA Jetson Nano, for instance, enables consumers to equip billions of low-power AI/ML systems with remarkable new features~\cite{suzen2020benchmark}. It paves the way for a wide variety of integrated IoT services, such as low-cost Network Video Recorders (NVRs), consumer automation, and analytics-rich gateways~\cite{basford2020performance}. With its ready-to-try applications and enthusiastic software developer community, Jetson Nano serves as the ideal tool for beginning students to gain knowledge about AI and robotics in real-life situations.

\item \textit{Cybersecurity}: Embedded systems are compact, specifically designed devices built to carry out a single task, frequently in real-time, while using as few resources as possible~\cite{de2019literature}. Installing protective measures on these platforms to guard against dangers like unauthorized usage or fraudulent attacks drives the need for embedded security~\cite{kumar2022securing}. These safeguards are included in electrical components, firmware, and applications to achieve an all-encompassing defence.
\end{enumerate*} 

\subsubsection{Networked Computing}
In this section, we discuss the main focus or paradigms, technologies or impact areas, and various trends or observations within networked computing.

\subsubsubsection{Focus/Paradigms} 
The following are the main focus or paradigms for networked computing:

\begin{enumerate*}[label =\color{black} \it \arabic*),itemjoin=\\\hspace*{\parindent}] 
\item \textit{Networking/Connectivity}: Servers in cloud computing underpin the services and Application Programming Interfaces (APIs) provided to internal and external clients. Communication on several levels, both inside and among data centers, is essential for effectively implementing cloud services~\cite{buyya2018manifesto}. Crucially, networking ensures that all parts can talk to one another in a safe, frictionless, effective, and adaptable way. 

Many developments and studies in networking during the past ten years have focused on the cloud~\cite{ren2019survey}. For instance, Software-Defined Networking (SDN) and Network Function Virtualization (NFV) aim to construct adaptable, versatile, and programmable computer networks to lessen the financial and time commitments of cloud service providers~\cite{wang2017integration}. Scalability challenges have spurred several current developments in network design for the Cloud Data Centers (CDCs), as well as the necessity for a flat addressing space, and the excess demand for machines. Notwithstanding these developments, numerous networking issues require a resolution. The excessive energy consumption of modern CDCs is a major issue~\cite{ren2019survey}. Especially because it is a common practice in data centers to have all networking equipment active at all times. 

Furthermore, unlike computing servers, most network parts (including switches, hubs, and routers), cannot be energy-proportionate; features like hibernation during periods of low traffic and connection-rate adaptability are not built in by default~\cite{ahmadabadi2023star}. Consequently, the design and execution of approaches and technologies that seek to minimize network energy usage and make it proportionate to the incoming load continue to be outstanding issues. 

QoS assurance presents another complex challenge within CDC networks~\cite{maray2022computation}. Service Level Agreements (SLAs) in modern clouds focus mostly on computing and storage. There is currently no way to encapsulate network performance constraints like latency and bandwidth assurances without resorting to ``best effort'' because no abstraction or method guarantees performance isolation. Providing network connections across widely dispersed resources (in other words, installing a ``virtual cluster'' encompassing resources in an amalgamated cloud setting), exacerbates this difficulty~\cite{bari2015orchestrating}. However, there are numerous open challenges to deliver reliability assurances for these networks---due to packages needing to navigate the (public) Internet, including resources in various locations~\cite{cai2022compute}.

\end{enumerate*}

\subsubsubsection{Technologies/Impact Areas} 
The key technologies and affected domains for networked computing include:

\subsubsubsubsection{Internet of Things (IoT)}
Devices (a.k.a., things) that can detect, control, and communicate are now routinely integrated with continuous control and monitoring functions via the Internet~\cite{buyya2018manifesto}. These devices have become ubiquitous in modern society, found in homes, on public transport, along highways, and in vehicles. Because of this, IoT applications may function in many contexts and provide a sophisticated evaluation and administration of complicated relationships~\cite{al2023energy}. As a result, IoT devices and services may solve problems in many application domains, such as digital health, facility administration systems, production, and transportation. IoT-based systems have to deal with limited processing power, memory, and storage space because (i) platforms are constantly changing, so devices that join a network have to be able to adapt to these changes; (ii) devices differ in how well they work with computers and what features they offer; and (iii) to ensure the safety of the IoT data that has been acquired, a federated system is needed~\cite{gubbi2013internet}. 

\textcolor{black}{These days, popular IoT use cases include medical care, smart cities, climate prediction, water supply management, and highway surveillance, all of which leverage the capabilities of cloud, serverless, fog, and edge computing for processing user data to meet QoS requirements~\cite{sriraghavendra2022dosp}.}

\begin{itemize}
\item \textbf{Healthcare:} \textcolor{black}{Among the many significant IoT applications is medical care, which is designed to treat conditions including heart attacks, diabetes, cancer, COVID-19, and influenza~\cite{verma2023fcmcps}. For instance, a patient's heart condition may be instantly diagnosed using a variety of medical devices in an interconnected IoT and computing environment~\cite{desai2022healthcloud}. Additionally, modern technology like Virtual Reality (VR) or AI can enhance the present healthcare system in the fight against inevitable pandemics~\cite{iftikhar2022fogdlearner}.}
\item \textbf{Agriculture: } \textcolor{black}{In order to forecast variables like yield, rainfall, and crop quality, the agricultural industry is making use of modern technology to analyze a wide range of data pertaining to agriculture~\cite{gill2017iot}. One use case is the development of cloud-based agricultural systems that can autonomously forecast the state of agriculture using data collected from a variety of IoT or edge sensors. Additionally, to facilitate automated farming, an iOS or Android application is created to handle the massive amounts of data and supply the information they need to the agriculturalists through their edge devices~\cite{sengupta2021mobile}.}
\item \textbf{Smart Home:} \textcolor{black}{Owners may optimize energy consumption and offer the necessary protection with the deployment of cameras through the implementation of smart homes, which allow them to operate their home devices from their cell phones~\cite{gill2019router}. For instance, a resource management approach that incorporates cloud and fog computing may be used for controlling edge devices utilizing a smartphone application, which in turn regulates the room's humidity, lighting, surveillance systems, fans, and voltage, such as via sensors connected to different household devices~\cite{iftikhar2022fog}.}
\item \textbf{Traffic Management}: \textcolor{black}{IoT is crucial in the efficient management of traffic through the use of a number of sensors and controllers~\cite{bansal2020deepbus}. To identify potholes, for instance, an IoT-based intelligent transportation system is created. In addition, its efficiency was assessed using a range of machine learning approaches and performance metrics~\cite{tuli2020ithermofog}. Additionally, data may be processed swiftly using fog and edge computing methodologies to notify about potholes early, thereby reducing the likelihood of mishaps.}
\item \textbf{Weather Forecasting:} \textcolor{black}{Through the use of cloud computing and the IoT, scientists and weather forecasters may better gather data to inform their work~\cite{singh2022quantum}. Scientists have long relied on visual observations, data storage, and the public presentation of meteorological factors like air quality and moisture to better understand and explain these phenomena~\cite{singh2021quantifying}. The findings may be made using an IoT system that relies on sensors and can transmit the results to the cloud.}
 \end{itemize}

Cloud services have long been relied upon by IoT applications to handle processing and permanent storage. Still, as the number of `things' proliferates, such services are increasingly unable to keep up with the real-time demands of IoT gadgets~\cite{stoyanova2020survey}. This is due to the high quantity of data and the short reaction times required by systems that operate in the real world over wide geographical areas. By moving resource orchestration from servers to edge networks, fog/edge computing expands the capabilities of cloud systems: Set up as a series of nested ``cloudlets'' that may perform data intake, processing, and administration~\cite{mansouri2020cloud}. Compared to cloud services, geographically localized solutions use less power and allow for more mobile resources by decreasing reaction times and increasing intake bandwidth through horizontal scalability. These features make fog/edge computing a potential future architecture for IoT applications since this architectural model allows for scalability on a logical and geographical scale with near-instantaneous response latency~\cite{othman2013survey}. 

By aggregating information from implanted and mobile gadgets and establishing mobile area networks, smart e-health apps can track information about patients in a continuous fashion~\cite{tuli2020healthfog}. By performing tasks like healthcare equipment noise filtering, data reduction and fusion, and analytics that identify harmful patterns in patients' well-being, smart gateways gather and interpret data from devices locally~\cite{gill2023chatgpt}. At the same time, longer-term patterns may be evaluated at cloud levels.

In addition, IoT systems supported by fog computing may adjust their actions based on the information they receive from sensors. For example, if a heart attack is recognized by initial processing at the fog layer, the intelligent gateway gathering signals from the defibrillator may adaptively boost the sample size before the attack. Similarly, the Industrial Internet of Things (IIoT) benefits from integrating edge, fog and cloud layers to provide specific and nearly real-time actions~\cite{vila2022edge}. Smart grids and energy management are central to the Internet of Energy (IoE) paradigm. Coarse-grained information on network health may be gathered from dispersed networks of energy producers that track power usage, generation and/or battery life. While `Smart-Meters' may communicate energy needs to service providers on a more detailed scale, monitoring capacity, generation, and use~\cite{tuli2020healthfog}. Therefore, IoT is a foundational technology for future systems, like electric automobiles and decentralized power grids~\cite{alahmad2021mobile}. In addition, the increased safety, reliability, and durability of electricity distribution that this type of grid may provide can better satisfy the evolving needs of consumers. In-depth surveys are a good resource for IoT researchers who want to explore more.
\\
\subsubsubsubsection{Software-Defined Network (SDN)}
SDN transcends traditional network paradigms by separating control logic from the underlying hardware and centralizing network management \cite{kreutz2014software}. This innovative approach facilitates programmable network architectures and streamlines management by distinctly segregating network policies, hardware implementation, and traffic forwarding~\cite{mekki2022software}. Integral to cloud computing, SDN enhances communication and automates configurations, revolutionizing network adaptability and resource utilization in diverse environments~\cite{son2018taxonomy}.  

NFV is another approach that utilizes software programs to perform traditionally hardware-based networking tasks, such as DNS, load balancing, and intrusion detection. NFV not only lowers costs but also enhances the flexibility of network functions and service responsiveness. Furthermore, VM consolidation in a virtualized network can help reduce energy costs by minimizing the number of VMs in operation~\cite{poutievski2022jupiter}. SDN-based cloud computing optimizes network virtualization while decreasing electricity consumption. Crucially, SDN increases the abstraction of physical assets and automates and optimizes the setup process~\cite{son2018taxonomy}. 

Many questions still need to be answered by scholars and investigators. First, ensuring data safety during transit across multiple cloud data centers is absolutely necessary for SDN-based cloud computing~\cite{kumar2022secure}. Second, even if SDN-enabled cloud infrastructures may be replicated, the balance between cost and energy use remains. Deploying SDN-based cloud computing systems is necessary to offer an economical network virtualization service with lower energy costs and greater dependability~\cite{kreutz2014software}. Furthermore, this may also boost data distribution and outcome collection by utilizing methods inspired by AI-based models, allowing us to expand existing information connectivity in such SDN contexts to accommodate blockchain-based systems.
%\end{enumerate*}

\subsubsubsection{Trends/Observations} 
The main trends and observations regarding networked computing are as follows: 

\subsubsubsubsection{Intelligent Edge}
The IoT connects billions of new devices, generating massive amounts of information that, inevitably, proves challenging to process. Over 41.6 billion IoT gadgets are estimated to be in operation by the end of 2025~\cite{wang2020convergence}. Increasing numbers of products, including connected autos, smart meters, and in-store sensors, are being created and installed by companies to improve customer experience while generating enormous quantities of data~\cite{gill2022ai}. 

Meanwhile, this emerging data must be gathered, managed, and processed immediately. \textit{What exactly will this mean?} Edge and fog computing might be a method for moving ahead. In the coming years, edge computing is forecast to receive far greater focus than fog computing. In contrast to traditional cloud computing, which analyzes data at a remote data center, edge computing performs so locally. In fog computing, it is possible to execute a portion of the work in the cloud, while edge devices perform other aspects~\cite{zhang2019mobile}. Since computing at the edge uses far less network bandwidth than conventional computing, the data exchanged among connected devices could take a long time. Computing it nearby, either on the gadget itself or within a local network, will be more cost- and energy-effective. On the contrary, edge computing may provide cloud computing with much-needed support in coping with the vast volumes of data created by the IoT and other connected devices~\cite{chen2019internet}. Emerging IoT devices create and transport data across the fog and edge, and their processing power is leveraged to carry out processes that could otherwise be performed in the cloud. Hence, managing these systems with fog and edge, IoT devices and support from the cloud requires distributing the intelligence along the computing tiers, which leads to edge intelligent~\cite{pujol2023edge}.

The terms ``fog'' and ``edge'' allude to these novel network nodes for IoT devices. Thus, they aid businesses in reducing their reliance on the cloud by transmitting information to analytics platforms. Businesses can lessen their dependency on cloud platforms for data processing and thereby reduce latency across networks by implementing edge and fog solutions~\cite{singh2023edge}. This will allow rapid evidence-based recommendations to assist them in their decision-making process. Nevertheless, once real-time processing is complete, edge devices must transfer data to the cloud for statistics to be performed on it~\cite{jia2020flowguard}. 

A company's communication network is largely concerned with enabling various remote apps and providing endless storage space, thanks to cloud computing, connectivity, and computing capacity. That will ultimately alter data processing at the edge in real-time, which is essential for optimal data utilization~\cite{chen2019internet}. Future-proof network infrastructures will need to accommodate an unprecedented influx of smart devices. For real-time intelligence, it is crucial to have the decision-making process located close to where the data is produced. Self-driving automobiles and self-sustainable, smart factory equipment, for instance, require to be making split-second decisions~\cite{yang2021edge}. Further, airline sensors collect data on engine efficiency in real time, allowing for predictive maintenance before a plane ever takes off. Potential cost reductions might be considerable. The more business connections an organization has, the more processing power and intelligence it can provide.

\subsubsubsubsection{6G and Beyond} 
The advancement to Industry 5.0 and the foundation of a technology-driven economy hinge on the development of Beyond 5G (B5G) and 6G networks. As communication and technological advancements increase, international industrial sectors will increasingly depend on 5G and B5G networks to provide revolutionary services and applications that will inevitably require ultra-low latency, unprecedented reliability, and continuous mobility~\cite{liu2022integrated}. Meanwhile, underpinned by Moore's law, mobile devices have been rapidly adopting systems-technology co-optimization (STCO) and related system-building approaches, which departs from the conventional system-on-a-chip (SoC) approach~\cite{ishtiaq2021edge}. 

Through cloud-based principles, including utilizing functioning between and among data centers, connecting in a micro-service setting, and concurrently offering reliable services and applications, it is expected that B5G/6G networks will be able to serve a wide variety of applications~\cite{kumar2021drone}. Both B5G and 6G networks aim to enable the smooth and complete integration of many industries, including the IoT, aerial networks (also known as drones), satellite accessibility, and submerged connectivity~\cite{shi2023machine}. To keep up with this astonishing expectation, the next generation of networks (B5G/6G) will largely rely on cutting-edge AI/ML technology for intelligent network operations and administration. B5G and 6G infrastructures are anticipated to provide computationally intensive applications and services paired with infrastructure shifts~\cite{alkhateeb2023real}. 

Edge computing has received a lot of interest and is being evaluated as an integrated service in 6G networks to enable the two fundamental changes in network infrastructure and network services. While many studies have focused on features like cache services and compute offloading methods, little is known about mobile edge computing implementation. The necessity of moving forward with a software-centric strategy from the network core to the wireless layer was emphasized in the first efforts that contributed significantly to the conception of 5G~\cite{ansar2023intelligent}. As with 5G networks, 6G networks will depend heavily on SDN, which, together with NFV, represents a departure from the conventional hardware-centric strategy~\cite{casamayor2023fundamental}. The mobile edge computing paradigm also encourages moving the base station (BS) and the core network functions to different places. BS functions are moved upstream to the cloud, and core network functions are moved downstream to the devices. The resulting boundary between the BS and the end device might be seen as an ``edge'' or ``fog'' domain~\cite{iftikhar2022fog}.

While cloud computing has made it possible for users to access richer and more complicated apps by tapping into the resources of a remote cloud server, an alternative technique is needed to meet the extremely delicate latency criteria stated for use cases in 5G and maybe 6G~\cite{akyildiz20206g}. This heterogeneous network design directly results from the complicated traffic distributions in today's wireless networks. A wireless access point (AP), a macro BS, and a small cell BS are just a few examples of network access nodes that may provide stable and smooth connections for mobile users~\cite{ghafouri2022mobile}. These network access nodes provide edge computing at network edges with less delay. The design of diverse mobile edge computing networks has gained more and more interest due to the varied properties of network access nodes, such as coverage capability and power transmission~\cite{akyildiz20206g}. Nevertheless, it is imperative to carefully plan for the offloading of computing tasks to many access nodes in a network~\cite{Wu2020Energy, Wu2023Lyapunov}. 

In a heterogeneous network design, intelligently distributing tasks and resources among different nodes can substantially boost system performance~\cite{basford2020performance}. For instance, by collaborating, the edge and cloud can elevate IoT tasks' QoS. While cloud servers manage compute-intensive tasks well, edge servers excel at processing tasks demanding minimal data or low latency~\cite{10335918}. Strategically assigning tasks among edge servers can redistribute work from overburdened nodes to less active ones, thus accelerating execution times.
 
\subsection{General Purpose vs. Specialized Computing}
Leveraging fit-for-purpose software and given enough time, general-purpose computing (which includes desktop PCs, laptops, mobile devices like tablets and smartphones, and even certain televisions) can handle just about any computation~\cite{owens2007survey}. A CPU, memory, input/output systems, and a bus are the main parts of any general-purpose computing system. In contrast, integrated computers, are used in intelligent systems, and are often referred to as ``special-purpose'' computing systems.

\subsubsection{General-Purpose Computing}
In this section, we discuss the main focus, paradigms, technologies, and impact areas, as well as various trends and observations about general-purpose computing. 

\subsubsubsection{Focus/Paradigms} 
The following are the main focus areas and paradigms associated with general-purpose computing:

\begin{enumerate*}[label =\color{black} \it \arabic*),itemjoin=\\\hspace*{\parindent}] 

\item \textit{Von Neumann Architecture}: 
A computing device with a Von Neumann architecture has its main components---the CPU, memory, and I/O---connected via a single bus~\cite{von2005john}. The efficiency of computers was greatly enhanced by the advent of this architecture, which provided effective means of storing and executing instructions. The fundamental idea behind this design is that data and instructions are handled in the same way. In other words, the data being handled and the program instructions themselves share the same storage and processing resources: a memory address can contain either an instruction to be executed or data; the software execution pathways decide how to interpret it~\cite{kimovski2023beyond}. This design substantially simplifies the framework and features of a computer, making it more accessible to both software engineers and non-technical users.

\item \textit{Integrated Computing}:
Compatibility throughout cloud applications and services is commonly achieved by implementing software adapters and libraries and deploying application containers for computing to facilitate mobility~\cite{yang2019integrated}. Nevertheless, there is still a variety of challenges that have existed since the beginning of cloud computing but, due to their complexities, have not been adequately resolved yet~\cite{ren2019survey}. One of these challenges is encouraging cloud connectivity without mandating a baseline set of capabilities for all services; ideally, customers can combine complicated features from several providers. Another area of investigation is how to develop cloud interoperability middleware that can facilitate the offering of complex services by composing more straightforward services from multiple 3rd-party providers~\cite{alsamhi2022computing}. Such a high degree of abstraction would empower users to make service decisions based on their requirements, such as price, turnaround time, privacy, etc. This brings up an additional key area that needs further study: the manner in which to allow user-level middleware (intercloud and hybrid clouds) to discover potential services for an ensemble without assistance from cloud service providers~\cite{chen2019deep}. A strategy that relies on cloud providers working together is unlikely to be successful because their financial goals lie in keeping all the features they offer to their consumers (i.e., they have no incentive to help due to the fact that just a portion of the offerings in an ensemble are themselves). Consequently, the middleware that allows the melody of services must address challenges at both of its connections: Firstly, the middleware should seamlessly and abstractly provide the service to cloud users. Secondly, for the consumers, a service might be implemented in its entirety by sub-services offered by one vendor (maybe leveraging a 3rd-part SaaS organization able to offer the functionality), or it might be acquired through composing multiple services from various providers~\cite{singh2021metaheuristics}. The provider user interface makes it possible to access complex functions without requiring special cooperation from providers~\cite{alsamhi2022computing}. The widespread use of cloud compatibility can offer commercial and financial advantages to cloud manufacturers, but frequently integrated clouds (which were achieved via Cloud Federation) cannot be realized until such time~\cite{botta2016integration}. This calls for the development of intercloud markets, distinctive approaches to invoicing and accounting, as well as novel cloud-suitable pricing systems.
 \end{enumerate*}

\subsubsubsection{Technologies/Impact Areas} 
The key technologies and affected domains for general-purpose computing include:

\subsubsubsubsection{Programming Models}
Clusters are a type of parallel or distributed computational system that consists of a group of interconnected standalone computers that work collectively as a single integrated computing resource. Clusters and grids are platforms that communicate with each other to serve as a single resource~\cite{cappello2005computing}. A multi-core parallel architecture describes this form of capability, which is based on specific functions. Conversely, cloud computing emerged on top of clusters to abstract leveraging their computing resources and coordinate enormous data sets. 

A programming model is tightly coupled to where data is transmitted to manage an application's functions. Important metrics to remember while building a programming model are efficiency, adaptability, goal architecture, and code maintainability~\cite{andrews2007achieving}. Data analytics software often handles massive data sets that require many phases of processing. Certain steps have to be carried out in order, while others are executed simultaneously across several nodes in a cluster, grid, or cloud. The capacity of algorithms to perform statistical analysis on huge amounts of data will be crucial to unlocking achievements in industrial advances and next-generation scientific discoveries~\cite{jackson2015survey}. 

With the exponential growth of data comes the difficulty of organizing massive data sets, which in turn increases their complexity due to the ways they connect with each other. Its many processes include moving, archiving, replicating, processing, and erasing data. Data life-cycle complexities can be reduced via solutions that automate and improve data management activities. It has been shown that two limitations affect the data life cycle~\cite{cao2018novel}. The framework used is the first limitation, initially regarding how it operates on data derived from consumers and apps. The second limitation derives from the observation that data is spread over several systems and infrastructures. That is why big data applications need to be capable of communicating amongst various systems that deal with the data and the effects that information and occurrences might have. The focus of this work is the second limitation, the big data infrastructure itself, and it includes a comprehensive analysis of the programming models and settings necessary to overcome this limitation. 

A programming model is underpinned by how quickly and smoothly its data is manipulated. A few elements to consider when creating a programming model include operation, adaptability, target designs, and the simplicity of maintenance code modification procedures~\cite{butts2007structural}. For the sake of service, it is sometimes necessary to sacrifice at least one of these aspects. The exchange of computation for data storage or transmission is a usual instance of algorithmic manipulation. These difficulties can be mitigated by employing parallel methods and technology. A software engineer might undoubtedly leverage many variants of the same technique to enable distinct performance adjustments on different hardware architectures. Modern computing clusters comprise nodes with more than one CPU, and their hardware designs range from tiny to super powerful.

\subsubsubsubsection{Virtualization}
With virtualization, the original physical object is replaced with a virtual one. The OSs of server infrastructure, hard drives, and PCs are some of the most typical targets for virtualization in a data center. Thus, virtualization decouples higher-level software and OSs from the underlying computing system~\cite{shen2021holistic}. 

VMs are a key component of hardware virtualization, standing in for a ``real'' computer running an OS. Emulating a computer system is what VMs do. A hypervisor makes a copy of the underlying hardware so that several OSs can share the same resources~\cite{jin2015h}. Despite being around for half a century, VMs are experiencing a surge in popularity because of the rise of the mobile workforce and desktop PCs. Server virtualization, which employs a hypervisor to effectively ``duplicate'' the underlying hardware, is a primary use case for virtualization technology in the corporate world~\cite{mansouri2021review}. In a non-virtualized setting, the guest OS generally works in tandem with the hardware~\cite{zhang2018performance}. 

OSs may be virtualized and continue functioning as if running on hardware, giving businesses access to similar performance levels while reducing hardware costs~\cite{alam2020survey}. The majority of guest OSs do not need full access to hardware; therefore, even if virtualization efficiency is lower than hardware efficacy, it remains preferred. This means firms are less reliant on a single piece of hardware and have more leeway to make necessary changes. 

Following the success of server virtualization, other sections of the data center have also begun to implement the same approach. Virtualization technology for OSs has been around for generations~\cite{xing2012virtualization}. In this implementation, the software enables the hardware to run several OSs in parallel. Companies that want to adopt a cloud-like IT infrastructure should prioritize virtualization. Using server resources more effectively is one of the primary benefits of virtualizing a data center~\cite{agache2020firecracker}. Thanks to virtualization, IT departments may use a single VM to host a wide variety of applications, workloads, and OSs, with the flexibility to add or subtract resources as required easily. The use of virtualization allows firms to expand readily. Organizations may better monitor resource utilization and react to shifting needs using such systems.

\subsubsubsubsection{Multicore Processors}
For improved performance and more efficient use of energy, integrated circuits with several processing cores, or ``cores,'' are becoming increasingly common. Furthermore, these processors enable more effective parallel processing and multithreading, allowing for simultaneous processing of numerous jobs~\cite{blake2009survey}. A computer with a dual-core arrangement is functionally equivalent to one with two or more individual CPUs. Sharing a socket between two CPUs accelerates communication between them. The use of processors with multiple cores is one technique to enhance processor performance while surpassing the practical restrictions of semiconductor manufacturing and design. Using several processors helps prevent any potentially dangerous overheating~\cite{gizopoulos2011architectures}. Multicore processors are compatible with any up-to-date computer hardware architectures. These days, multicore processors are standard in desktop and portable computers. Nevertheless, the actual power and utility of these CPUs depend on software built to leverage parallelism~\cite{delgado2020new}. Application tasks are broken up into many processing threads in a parallel strategy, distributing and managing them over multiple CPUs.
%\end{itemize}

\subsubsubsection{Trends/Observations} 
The main trends and observations regarding general-purpose computing are as follows:

\subsubsubsubsection{Software Systems}
Web-based computing and Software Engineering (SE) are closely related disciplines. For instance, service-oriented SE provides various advantages to the software creation procedure and app development by merging the greatest elements of services and the cloud. In contrast to cloud computing, which is concerned with effectively transmitting services to consumers using adaptable virtualization of resources and load balancing, service-oriented SE is concerned with architectural design (service searching and composition)~\cite{piattini2021toward}. 

Customers and developers are both essential to the evolution of hardware innovations, which is why software engineering is a crucial discipline~\cite{arvanitou2021software}. With the help of distributed computing and virtualization, customers may set up automatically managed VMs and cloud services for their initiatives and applications. Thanks to cloud services, teams working on software may now more easily collaborate on the development, testing, and distribution of their products. Here are some scenarios in which cloud computing might improve software engineering: The production timeline can be compressed~\cite{althar2021realist}. As a result of the availability of ample computing resources made possible by cloud computing and virtualization, software engineers no longer need to rely just on a single physical computer. The time it takes to install the required applications may be decreased by retrieving cloud services, indicating that development activities can be performed with increased parallelism thanks to cloud computing. Third, VMs and cloud instances may substantially improve the setup and delivery procedures. 

Using sufficient virtualization resources from a private or public cloud, developers can speed up the building and testing process, which is otherwise, extremely time-consuming~\cite{xing2012virtualization}. To circumvent this, a simplified system for managing code versions is required. In software development, code branches are used for refining and adding features. With cloud computing, there's no need to invest in or lease expensive hardware only to store some code. A distributed software engineering team may access apps more easily in a cloud setting, and service quality can be enhanced through dynamic resource allocation. As a result, the software construction process is streamlined thanks to cloud computing, which eliminates the need for development servers to rely on specific physical computers~\cite{arvanitou2021software}. Nevertheless, there are obstacles when merging software engineering and cloud computing. The majority of the difficulties are with moving the data. Because different cloud providers use various APIs to offer cloud services, migrating software and data from one cloud to another while avoiding vendor lock-in is challenging. Avoiding over-reliance on any one set of APIs is one way to fix this problem while building and releasing applications in the cloud. The problem of dependability and accessibility is another obstacle. If everything is moved to the cloud, it will be difficult to retrieve the data if the cloud is compromised by hackers or affected by an unexpected calamity. The engineering teams are responsible for creating a local backup of their work~\cite{de2022software}. 

Cloud computing allows software engineering academics to study multinational software development. Several investigations have examined the feasibility of using cloud computing to lower operational, delivery, and software development expenses. Researchers have investigated the feasibility of replacing services with a cloud-based platform for student-to-student knowledge exchange and collaboration~\cite{sharma2022applications}. Software systems have been supplanted by systems running on the cloud to reduce expenditures and maximize the utilization of resources. The conventional data management techniques have become increasingly cumbersome in the past few years due to the rapid increase in available data. A new frontier for study in software engineering has opened up thanks to the IoT, Blockchain (the distributed ledger), and ML/AI, with data management being the primary challenge~\cite{al2019blockchain}. These studies also provide a springboard for further study and innovative approaches to cloud data management, leading to the development of advanced technologies like Cisco's pioneering fog computing~\cite{doyle2022blockchainbus}. Enterprise software developers are creating an abstraction layer, or ``Blockchain-as-a-Service'', and selling it to other businesses as a subscription service~\cite{gill2022ai}. These numerous new fields rely significantly on software engineering, yet they could not exist without it.

\subsubsubsubsection{Simulations}
The capacity to carry out research, analyze strengths and shortcomings, and demonstrate viability is hampered in new or emerging computing domains due to the lack of mature technology and sufficient infrastructure. In many cases, the time and resources needed to acquire the necessary physical resources make it impractical to conduct the necessary research~\cite{mansouri2020cloud}. An alternative approach that can approximate a physical environment is a simulator. Additionally, simulation offers the ability to test suggested hypotheses in lightweight and low-cost settings. Real-world testing of novel methods is difficult and expensive because of the time and effort required to gather the necessary hardware resources (particularly for large-scale tests) and create the necessary software applications and systems~\cite{jurado2021simulation}. Investigators demonstrate the viability of their ideas by modelling and simulation, and then conduct tests to confirm their concepts in a monitored environment utilizing simulation tools. Simulation software provides a convenient setting for testing solutions to real-world issues by allowing users to experiment and see what happens~\cite{buyya2022strategy}. If a commercially available simulator is inadequate for user needs, then researchers should consider building their own, complete with graphical user interfaces. This is especially true if users need to simulate components of emerging computer architectures~\cite{brady2016all}. Researchers could benefit greatly from using a simulator to formulate questions and analyze different theoretical frameworks in simulated setups, therefore stimulating more research and fostering the development of communities within the relevant field.
%\end{itemize}

\subsubsection{Specialized Computing}
In this section, we discuss the main focus or paradigms, technologies or impact areas, and various trends or observations within specialized computing. 

\subsubsubsection{Focus/Paradigms} 
The following are the main focus or paradigms for Specialized Computing:

\subsubsubsubsection{Reconfigurable Computing}
The modern paradigm of reconfigurable computing enables hardware components to swiftly alter their configuration and functioning in response to changing processing needs. Reconfigurable computer devices, such as Field-Programmable Gate Arrays (FPGAs), can be reprogrammed to perform a variety of different functions~\cite{ferraz2021survey}. The main function of reconfigurable computing is to fill the void among general-purpose CPUs and Application-Specific Integrated Circuits (ASICs)~\cite{compton2002reconfigurable}. It allows hardware to be optimized for efficiency, power efficiency, and flexibility by matching application requirements. Static and dynamic switching are the two primary modes of operation for reconfigurable technology. In static reconfiguration, the component settings are adjusted prior to the computer starting to compute. However, dynamic reconfiguration permits hardware changes to be made while the system is running, allowing for dynamic modifications to hardware behavior.

\subsubsubsubsection{Domain-Specific Architectures}
As computing and the digital transformation spread to various use cases, such as cloud (AI/HPC), networking, edge, the IoT, and self-driving cars, highly domain-specific computational tasks are making it more likely that Domain-Specific Architectures (DSAs) can enable big performance gains~\cite{jouppi2018domain}. Using ChatGPT and other comparable software that are powered by large language models--—which are fundamental to achieving generative AI--—provides greater specialization inside AI workloads at extremely high volume, which motivates further hardware specialization~\cite{gill2023chatgpt}. DSAs, or application-domain-specific hardware and software, have substantial market potential. As a result of their superior performance on tasks that profit from a significant amount of parallel computing, such as AI workloads (learning and predicting), GPUs and Tensor Processing Units (TPUs) are currently controlling a sizable portion of the data center market~\cite{cong2018customizable}. Meanwhile, accelerations of 15–50 times the original speed, depending on the workload, are not uncommon. In the automobile industry, minimal latency and high-performance inference are provided via tailor-made solutions from industry leaders.

\subsubsubsubsection{Exascale Computing}
To handle the massive computations required by convergent modelling, simulation, AI, and data analysis, an entirely novel type of supercomputer called exascale computing has emerged~\cite{lindsay2021evolution}. This is motivated by advanced computational needs in science and engineering. 

Exascale computing (also supercomputing) becomes essential to expedite the generation of knowledge. Researchers and technologists may employ data analysis driven by exascale supercomputing to expand the frontiers of our existing understanding and promote breakthrough ideas. Supercomputing capabilities are in high demand as the world moves towards exascale computing to ensure continued scientific and technological advancements, while our civilization's technological and scientific frameworks are progressing quickly thanks to exascale computing~\cite{ji2022magnetic}. The immense potential of these tools necessitates their careful operation, especially as cultures worldwide undergo rapid changes in their moral frameworks and their perceptions of what it means to live sustainably. As such, novel responses to formerly intractable issues are being uncovered thanks to exascale computing. 

Exascale supercomputers are prohibitively expensive to construct; thus academics and scientists rely on funding to lease them instead of buying their own~\cite{heldens2020landscape}. Exascale computing systems produce enormous quantities of heat because of the volume of data they process. They require extremely cold environments to be stored in or unique cooling mechanisms built into the systems and racks themselves for optimal performance. Differentiating them from other types of supercomputers and quantum computers, they are computer systems with the largest capacity and most powerful hardware~\cite{kim2023evidence}. 

To further our understanding of the universe, exascale computers can model elementary physical processes like the granular interactions of atoms. Quite a few sectors rely on this capacity to analyze, forecast, and construct the world of tomorrow: for instance, better predict the weather, investigate in detail the interaction between rain, wind, clouds, and various other atmospheric occurrences, analyze their effects on one another at a molecular level and so on. Mathematical formulas can be used to determine the millisecond-by-millisecond effects of all forces acting in a certain environment at a specific time~\cite{anzt2020preparing}. These seemingly trivial interactions rapidly generate billions of possible permutations, which need trillions of mathematical equations to calculate and analyze. This kind of speed is only achievable on an exascale machine. By studying the results of these computations, researchers can gain a deeper insight into the nature of our universe~\cite{kim2023evidence}. Exascale supercomputers, despite their challenges, can literally increase our understanding, enabling us to address the problems of the future.

\subsubsubsubsection{Analog Computing}
A novel approach may minimize errors in ultra-fast analog optical neural networks. Larger and more complicated machine-learning models need stronger and more effective computing gear. However, standard digital computers are lagging. Compared to a digital neural network, an analog optical network's performance in areas like image classification and voice recognition is comparable. However, its speed and energy efficiency far exceed those of its digital counterparts~\cite{zangeneh2021analogue}. Nevertheless, hardware faults in these analog devices might impact the accuracy of calculations. One possible source of this inaccuracy is microscopic flaws in the hardware itself~\cite{zhang2020neuro}. Errors tend to multiply rapidly in a complex optical neural network. Even when using error-correction approaches, due to the basic features of the components that make up an optical neural network, a certain degree of error is inescapable~\cite{zhao2020reliability}.  Conversely, the optical switches that make up the network's architecture can reduce mistakes they typically accrue by adding a modest hardware component.

\subsubsubsubsection{Neuromorphic Computing}
When applied to AI, neuromorphic computing makes it possible for AI to learn and make decisions independently, significantly improving over the first generation of developing AI. To acquire abilities in areas like recognizing voice and sophisticated tactical games, including chess and Go, neuromorphic algorithms are now involved in deep learning~\cite{zangeneh2021analogue}. 

Next-generation AI will imitate the human brain's capacity to comprehend and react to circumstances instead of merely operating from formulaic algorithms~\cite{zador2023catalyzing}. When it comes to understanding what they've read, neuromorphic computing systems will seek out patterns and use their `common sense' and the surrounding context. When Google's Deep Dream AI was programmed to hunt for dog faces, it notably showed the limitations of algorithm-only computer systems~\cite{zhao2020reliability}: Any images that it interpreted as having dog faces were transformed into dog faces.

Third-generation AI computing attempts to simulate the elaborate structure of a living brain's neural network~\cite{schuman2022opportunities}. This calls for AI with computing and analytic capabilities on par with the extremely efficient biological brain. To demonstrate their exceptional energy economy, human brains can surpass supercomputers using less than 20 watts of electricity. Spiking Neural Networks (SNN) are the AI equivalent of our synaptic neural network~\cite{morabito2024advances}. They leverage many layers of artificial neurons, and each spiking neuron may fire and interact with its neighbors in response to external inputs. 

Most AI neural network architectures follow the Von Neumann design~\cite{von2005john}, which divides the memory and computation into discrete nodes. Computers exchange information by reading it from memory, sending it to the CPU for processing, and then returning it to storage. This constant back-and-forth wastes a lot of time and effort. It causes a slowdown that becomes more noticeable while processing huge data sets. As a response, multiple neuromorphic devices can be utilized to supplement and improve the performance of traditional technologies, such as CPUs, GPUs, and FPGAs~\cite{zhang2020neuro}. Low-power neurological systems may perform powerful activities, including learning, browsing, and monitoring. A practical instance would involve immediate voice recognition on mobile phones without the CPU needing to interact with the cloud.
%\end{itemize}

\subsubsubsection{Technologies/Impact Areas} 
The key technologies and affected domains for Specialized Computing include:

\begin{enumerate*}[label =\color{black} \it \arabic*),itemjoin=\\\hspace*{\parindent}] 
\item \textit{Graphics Processing Unit (GPU)}: GPUs have rapidly risen in prominence as a crucial component of both home and enterprise computers~\cite{owens2008gpu}. A GPU is a special type of computer chip deployed in a variety of application domains, most notably the rendering of moving images. While GPUs are best recognized for their usage in gaming, they are also finding increasing application in the fields of creative creation and AI~\cite{rosenfeld2022query}. The initial purpose of GPUs was to speed up the display of 3D visuals. They improved their functionality as they got more adaptable and programmable over time. This paved the way for more complex lighting and shadow characteristics and photorealistic environments to be implemented by graphics developers. Additional engineers started using GPUs to drastically speed up various tasks in deep learning, HPC, and other fields~\cite{ferraz2021survey}.

\item \textit {Compute Unified Device Architecture (CUDA)}: The demand for more powerful computers grows daily. As a result of constraints like size, climate, etc., vendors throughout the world are finding it difficult to make future improvements to CPUs~\cite{owens2008gpu}. Service providers that provide solutions in this kind of environment have begun to seek out performance improvements elsewhere. The use of GPUs for parallel processing is one option that enables significant speed gains~\cite{sanders2010cuda}. The total number of cores in a GPU is significantly greater than that of a CPU. Although CPUs are designed for sequential processing, offloading them to GPUs enables parallel processing. For general-purpose computing on NVIDIA's GPUs, users can rely on CUDA, which allows for the execution of processes in parallel on the GPU without any specific order requirement~\cite{ferraz2021survey}. Offloading compute-intensive activities to Nvidia's GPU using CUDA is straightforward thanks to the library's support for the popular C, C++, and Fortran programming languages~\cite{sanders2010cuda}. CUDA is employed in scenarios needing extensive computational power or suitable for parallel processing to achieve high performance. Fields such as AI, healthcare analysis, science, digital transformation, cryptocurrency mining, and scientific modeling, among others, depend on CUDA technology.
\end{enumerate*}

\subsubsubsection{Trends/Observations} 
The main trends and observations regarding Specialized Computing are as follows:

\textit{Large-Scale ML}: As big data grows, ML algorithms with many variables are needed to ensure that these models can handle very large data sets and make accurate predictions, including hidden features with many dimensions, middle representations, and selection functions~\cite{tuli2020predicting}. The need for ML systems to train complicated models with millions to trillions of variables has increased as a result~\cite{lwakatare2020large}. Distributed clusters of tens to hundreds of devices are often used for ML systems because they can handle the high computing needs of ML algorithms at these sizes. Yet, developing algorithms and software systems for these distributed clusters requires intensive analysis and design~\cite{wang2020survey}. The latest advances in industrial-scale ML have focused on exploring new concepts and approaches for (a) highly specialized monolithic concepts for large-scale straight applications, such as different distributed topic models or regression models, and (b) for adaptable and readily programmable universally applicable distributed ML platforms such as GraphLab based on vertex programming and Petuum using a parameter-driven server~\cite{angenent2020large}. It is widely acknowledged that knowledge of distributed system topologies and programming is essential; however, ML-rooted statistical and algorithmic discoveries can yield even more fruit for large-scale ML systems in the form of principles and techniques specific to distributed machine learning applications. These guidelines and techniques shed light on several crucial questions:
\begin{itemize}  
\item How to share an ML application among nodes? 
\item How to connect machine-learning calculations with machine-to-machine dialog? 
\item How should one proceed with having such a conversation? 
\item What ought to be conveyed among machines? And, should they cover many big ML-related topics, from practical use cases to technical implementations to theoretical investigations~\cite{shi2023machine}? 
\end{itemize}  
Understanding how these concepts and tactics may be made effective, generally applicable, and easy to develop is the primary goal of large-scale ML systems studies, as is ensuring that scientifically validated accuracy and scalability assurances underpin them.
%\end{itemize}

\subsection{Centralized vs. Decentralized Computing}
A central server controls and processes most of the data in a centralized network, whereas no single entity has influence over a decentralized network.
\subsubsection{Centralized Computing}
In this section, we discuss the main focus or paradigms, technologies or impact areas, and various trends or observations within centralized computing.  

\subsubsubsection{Focus/Paradigms} 
The following are the main focus or paradigms for centralized computing:

\begin{enumerate*}[label =\color{black} \it \arabic*),itemjoin=\\\hspace*{\parindent}] 
\item \textit{Cloud Computing}: The adoption of cloud computing, which revolutionized how end-users and software engineers interact with applications and computing systems, led to the rise of technology as the fifth utility~\cite{buyya2018manifesto}. Cloud computing was successfully accepted by giving consumers on-demand access to the computing power they want, the freedom to modify their resource consumption as needed, and the transparency of paying just whatever is being utilized. Business groups, regulatory bodies, and universities have all been quick to endorse it since it first appeared. Like contemporary society relying on essential utilities, the cloud has grown into the economy's foundation by providing immediate utilization of subscription-driven computing resources~\cite{buyya2009cloud}. As a result of using cloud technology, innovative companies can be launched quickly, existing ones can expand globally, advances in science can be sped up, and novel computing methods can be developed for ubiquitous and pervasive apps~\cite{anwar2021recommender}. SaaS, PaaS, and IaaS have served as the three primary service models that have pushed uptake in the cloud thus far~\cite{durao2014systematic}.
\end{enumerate*}
\begin{itemize}
\item \textit{Mobile Cloud Computing}: To provide value to mobility consumers, network operators, and cloud service providers, mobile cloud computing integrates mobile devices, cloud computing, and communication networks. With the help of mobile cloud computing, a wide variety of handheld gadgets can run complex mobile apps. Under this paradigm, handling and storing data is done by servers rather than individual mobile devices~\cite{othman2013survey}. Several advantages result from the use of mobile cloud computing apps based on this architecture: (i) battery life has significantly increased; (ii) there has been an increase in both the speed and size of data being stored and processed; (iii) the system's emphasis on ``store once, access anywhere'' eliminates complex data synchronization; and (iv) stability and scalability have been dramatically enhanced. Nevertheless, inadequate network capacity is a significant challenge for mobile cloud computing~\cite{alahmad2021mobile}. Wireless mobile cloud services have capacity constraints in contrast to their cable counterparts. The spectrum of mobile devices offers a wide range of wavelengths. This has resulted in slower access speeds, as much as one-third in comparison to a wired network. Due to the increased likelihood of data loss on a wireless network, it is more challenging to recognize and deal with security risks on mobile devices than on desktop computers~\cite{malik2021effort}. Customers frequently report issues with accessibility to services, including network outages, overcrowding on public transit, lack of coverage, etc. Customers may occasionally experience a low-energy signal, which slows down access and impacts data storage. Mobile cloud computing is employed on several OS-driven platforms, including Apple iOS, Android, and Windows Phone, resulting in network modifications that need cross-platform compatibility~\cite{jin2022survey}. Mobile gadgets have a greater environmental impact due to their high energy consumption and low output~\cite{ren2019survey}. As the use of mobile cloud computing grows, so does the problem of the increased drain on mobile devices' batteries. A device's battery life is crucial for using its software and executing other tasks. Although the modified code is tiny in size, offloading uses more energy than running it locally~\cite{patros2021toward}.
\item \textit{Green Cloud Computing}: In the last few decades, Information and Communication Technology (ICT) has significantly evolved, drawing on technological advancements from the past two centuries. This evolution has elevated computing to the status of a fundamental service, akin to traditional utilities such as water, electricity, gas, and telephony, thereby establishing it as the fifth essential utility in modern society~\cite{masdari2020green}. Modern cloud computing systems are becoming progressively large-scale and dispersed as more and more businesses and organizations have shifted their computing workload to the cloud---while others opt out of maintaining code altogether and instead leverage cloud-powered SaaS services. A cloud computing infrastructure of this magnitude not only offers more affordable and dependable services but also, increases energy effectiveness and reduces the global community's carbon impact~\cite{gill2018taxonomy}. Every minor enhancement is much appreciated. In an effort to achieve zero carbon emissions, the community has recently been aggressively exploring a more sustainable version of cloud computing called green cloud computing to lessen reliance on fossil fuels and curb its carbon footprint~\cite{shu2021research}.

Green cloud computing is a system that considers its constraints and goals to minimize energy consumption. Researchers are focusing on scheduling workloads and resources in light of carbon emissions, in order to increase the effectiveness of the resources used~\cite{zhou2020energy}. Additionally, forecasting problems with hardware and creating management systems to use hardware with varying degrees of dependability can maximize device lifetime and reuse. Further, utilizing micro-data centers---rather than standard server data centers---is a promising approach to boost efficiency and save costs. These facilities can accommodate future growth, serve huge populations, and dissipate heat effectively~\cite{mansour2023design}. 

Furthermore, virtualization is another ecologically friendly technique that boosts the versatility of system resources. Through improved tracking and control, servers may pool their resources more effectively~\cite{singh2021dynamic}. Innovations and practices that support sustainable development are constantly being developed as organisations rely more heavily on cloud services to enable ``green cloud computing''.
\end{itemize}

\subsubsubsection{Technologies/Impact Areas} 
The key technologies and affected domains for centralized computing include:

\begin{enumerate*}[label =\color{black} \it \arabic*),itemjoin=\\\hspace*{\parindent}] 
\item \textit{Cloud Storage Technologies}: Files and information stored in the cloud may be accessed from anywhere with a web connection or via secure network access. Transferring files to the cloud puts the responsibility for data security squarely on the shoulders of the cloud provider, rather than consumers. The service provider hosts, manages, and maintains the servers where user data is stored, and they also guarantee that users always have access to their files~\cite{zeng2009research}. When compared to storing data on local discs or storage networks, cloud-based storage is a more affordable and scalable option. There is a limit to the quantity of information that can be stored on a hard disc. When users exhaust internal storage space, they must copy their data to removable media. The difference between on-premises storage networks and cloud storage is that the latter sends data to servers located in a remote data centre. VMs, which are abstracted on top of an actual server, make up the vast majority of users' servers~\cite{hota2023leveraging}. Known as autoscaling, a cloud provider spins up more virtual servers as necessary to accommodate users' ever-increasing storage demands. Files, blocks, and objects are the three primary categories of cloud storage, which are accessible in private, public, and hybrid cloud configurations.
\item \textit{Microservices}: Microservices are a type of application architecture in which several autonomous services collaborate using simple APIs. A cloud-native software development method, microservice architecture separates an app's main functionality into its own modules~\cite{kumar2023qos}. By compartmentalizing the app's components, the development and operations teams may collaborate without interfering with each other. If several engineers can collaborate on the same project simultaneously, it takes less time to complete. This is in contrast with the monolith software architecture, which had been the standard for application development in the past~\cite{ghofrani2018challenges}. 

All of an app's features and services are tightly bound together and run as one seamless whole under a monolithic architecture~\cite{song2023chainsformer}. The application's architecture becomes more involved whenever new features are introduced, or existing ones enhanced. Because of this, optimizing a single feature inside the application requires disassembling the whole thing, which is a time-consuming and tedious process. This additionally necessitates that scaling the application as a whole is required if scaling any one process inside it---rather than just scaling out just that overloaded element~\cite{al2022ai}. 

Microservice architectures separate an app's essential features into individual processes. To adapt to shifting business requirements, software engineering teams may develop and maintain new elements independently of the rest of the application. The monolith has been the standard for application development in the past. An application's features and services are tightly bound together and run seamlessly under a monolithic architecture~\cite{xu2022coscal}.

Microservices' malleability might hasten the deployment of novel modifications, necessitating the development of novel patterns. In software engineering, a ``pattern'' is supposed to refer to any mathematical approach that is known to function. An ``anti-pattern'' is an erroneous pattern that is often applied to achieve a solution but often ends up causing even more problems.
\item \textit{Container Technologies}: Given the advent of Docker, container technology has gained widespread use in the cloud computing sector, where it is used to efficiently execute user workloads~\cite{bentaleb2022containerization}. Since containers are independent entities that may run without sharing data with other containers, this technology provides an inexpensive cloud environment for deploying applications. In a container, applications deployed on the same hardware server can share the same underlying resources while maintaining their own distinct processes~\cite{barbalace2020edge}. 

Container technology leverages Linux kernel capabilities, such as \texttt{libcontainer} and control groups (\texttt{cgroups}). By utilizing \texttt{cgroups} and namespaces, Docker can operate containers independently within a host node, providing the container with its own dedicated set of runtime resources (including the host's networked devices, disc space, memory, and CPU). In addition, namespaces provide for more efficient application deployment and development by separating the program's perspective from the operating environment~\cite{golec2020biosec}. Furthermore, containerization becomes an example of creating, publishing, and running applications in an isolated way and is indicated as a Container as a Service (CaaS). There are three primary advantages of CaaS: (1) containers boot up in no time at all; (2) they consume fewer resources than VMs; and (3) many instances may be operated at once using container technology~\cite{struhar2020real}. 

Recent investigations~\cite{casalicchio2020state} into container technology reveal unanswered research questions. Firstly, containers are less secure than VMs since they share the kernel, but this is something that may be fixed in future versions with the help of Unikernel. Secondly, optimizing container performance is a time-consuming endeavor that requires buffer space. Swarm and Kubernetes are two examples of cutting-edge cloud computing tools that may be used for handling user-created QoS-based container clusters~\cite{zhong2020cost, mallikarjunaradhya2023overview}. Thirdly, because containers share the same computing/hardware resources, co-located tenants can suffer from unpredictable performance interference when the CPU Shares algorithm is used, and even worse, they can leak information enabling side-channel attacks to be performed by a malicious tenant~\cite{patros2016investigating}.

\item \textit{Serverless Computing}: The use of serverless computing in the creation of apps for the cloud is gaining traction~\cite{kounev2021toward}. The goal of serverless computing is to ensure that only the most effective serverless technologies are deployed, reducing costs while increasing benefits~\cite{shafiei2022serverless}. Meanwhile, companies in all industries are adopting AI since it is the next generation of innovation. Due to these AI-driven platforms, we've been able to make more accurate, timely decisions~\cite{golec2023qos}. They have altered the methods used to conduct business, communicate with customers, and assess company information. Complex ML systems can significantly affect developers' output and efficiency~\cite{aslanpour2021serverless}. However, switching to a serverless architecture may be able to solve many of the issues that engineers face. The serverless design ensures that the machine learning models are administered correctly and that all available resources are utilized efficiently. Developers will be able to devote a greater amount of time to training AI models rather than maintaining the server environment~\cite{li2022serverless}. Creating ML algorithms is a common practice when confronting difficult problems. They perform tasks such as data analysis and preprocessing, model training, and AI model tuning~\cite{li2022serverless}. Serverless computing running AI tasks will provide for reliable data storage and communication.
\end{enumerate*}

\subsubsubsection{Trends/Observations} 
The main trends and observations regarding centralized computing are as follows:

\begin{enumerate*}[label =\color{black} \it \arabic*),itemjoin=\\\hspace*{\parindent}] 
\item \textit{AI-driven Computing}: The fundamental benefit of autonomic computing is a reduced overall cost of ownership. Therefore, the cost of upkeep will be drastically reduced. The number of technicians required to keep everything running smoothly will go down as a result, too. Autonomous IT systems driven by AI will reduce the time and money needed for installation and upkeep while also improving IT system stability~\cite{gill2022ai}. 

In accordance with higher-order benefits, businesses would be more capable of handling their operations with the help of IT systems that are able to adopt and execute directions based on their business plan and allow for adjustments in reaction to evolving circumstances. Using AI-based autonomic computing has several advantages, including reducing the expense and quantity of human labor needed to manage large server farms, which is made possible through server consolidation~\cite{kumar2023ai}. Using AI for self-driving computers will simplify system administration. As a result, computer systems will be greatly enhanced. Server load distribution is another potential use case since it allows for parallel data processing across several computers. Meanwhile from an energy perspective, analyzing the power grid in real-time allows for more cost-effective and long-term power policy decisions to be made~\cite{buyya2018manifesto}. 

There are benefits to using remote data centers instead of keeping data in-house. Despite the hefty upfront expenses, businesses may obtain AI technology relatively easily by paying a monthly fee on the cloud. When employing an AI-powered system, there may be no need for human involvement in data analysis~\cite{iftikhar2022tesco}. Using AI in the cloud can potentially make businesses more effective, strategic, and insight-driven. AI can increase output by automating routine processes and data analysis without human intervention~\cite{bansal2020deepbus}. For instance, integrating AI technology with Google Cloud Stream statistics could enable real-time personalization, anomaly detection, and management scenario prediction~\cite{firouzi2022convergence}. As the number of cloud-based applications grows, it is essential to implement a system of rigorous data protection based on intelligence. Network security systems backed by AI-enabled traffic tracing and analysis; AI-enabled devices can sound an alarm as soon as an anomaly is detected. Such methods will ensure keeping sensitive data protected.

\item{Net Zero Emissions}: Several data center operators have committed to being carbon neutral by the year 2030 as sustainability becomes an increasingly hot subject in the industry~\cite{cao2022toward}. But are these promises only a reaction to the possibility of legislation, or is it actually making progress? If business planes are a major contributor to global warming, how do they plan to cut their carbon footprint so rapidly? The data centers' businesses in the United States use about as much power as the state of New Jersey~\cite{siddik2021environmental}. If all of the power came from renewable resources, this level of demand would not be a problem. Liquid cooling and energy generation both require water, and a typical data centre uses as much water as an urban area of 30,000 to 50,000 individuals~\cite{senthilkumar2023enhancement}. Becoming a pioneer in sustainability might also bring up emerging markets. Companies are going to utilize green data centers to offset their carbon footprints as they grow and become more energy efficient and sustainable~\cite{kurniawan2023decarbonization}. A car company, for instance, might employ emission-free data centers for all of its corporate services. Last but not least, adopting environmentally friendly practices may help businesses comply with environmental rules, avoid fines, and get access to attractive, low-interest, long-term capital investment possibilities~\cite{wilkinson2024environmental}.            

\end{enumerate*}

\subsubsection{Decentralized Computing}
In this section, we discuss the main focus or paradigms, technologies or impact areas, and various trends or observations within decentralized computing.

\subsubsubsection{Focus/Paradigms} 
The following are the main focus or paradigms for decentralized computing:

\begin{enumerate*}[label =\color{black} \it \arabic*),itemjoin=\\\hspace*{\parindent}] 
\item \textit{Parallel Computing}: Through the utilization of several processor cores, parallel computing can perform multiple tasks simultaneously. The ability to divide and conquer a work into smaller, more manageable chunks is what sets parallel computing apart from its serial counterpart~\cite{bhardwaj2020heart}. Real-world events may be modelled and simulated effectively on parallel computing systems \cite{fox2014parallel}. As processing and network speeds continue to increase at an exponential rate, adopting a parallel architecture is no longer just a nice-to-have. The IoT and big data will eventually require us to process terabytes of data simultaneously. Devices such as dual-core, quad-core, eight-core, and even 56-core CPUs utilize parallel computing. Therefore, although parallel computers are not brand new, this is the problem: These new technologies are spitting up ever-faster networks, and computer efficiency has surged 250,000 times in 20 years~\cite{wu2021multi}. For instance, AI technologies will sift through more than 100 million patients' heart rhythms in the medical sector alone, looking for signals of A-fib or V-tach, saving many lives in the process~\cite{fox2014parallel}. When the systems must slowly move through each procedure, they will not be able to complete it on time. As great as the potential is, parallel computing may be nearing the edge of what it can achieve with conventional processors. Parallel calculations may see significant improvements in the coming decade, thanks to quantum computers. In a current, unauthorized announcement, Google claimed to have achieved \textit{quantum supremacy}~\cite{gill2021quantum, singh2022quantum}. If it is accurate, then Google has created a machine that can perform in 4 minutes whatever would require the most capable supercomputer on the planet 10,000 years to achieve~\cite{gill2022quantum}. Quantum computing is a major step forward for parallel computation. Imagine it like this: Processing in a serial fashion does one task at a time. An 8-core simultaneous computer can do eight tasks simultaneously. There are fewer particles in the universe than there are qubits' states in a 300-qubit quantum computer~\cite{gill2021quantum}.
\item \textit{Fog Computing}: The proliferation of IoT devices and the effort needed for analyzing and storing enormous amounts of knowledge led to the development of fog computing as a complementary service to traditional cloud computing. Fog computing, which provides fundamental network functions, can back IoT apps that require a small response-time window~\cite{iftikhar2022ai}. Due to the dispersed, diverse, and constrained nature of the fog computing paradigm, it is challenging to spread IoT application operations effectively within fog nodes to meet QoS and Quality of Experience (QoE) limitations~\cite{singh2021fog}. Vehicle-to-Everything (V2X), medical tracking, and manufacturing automation adopt fog computing as it delivers the ability to compute close to the consumer to match fast response demands for these applications. Due to the proliferation of IoT devices, these applications generate massive volumes of data. Cloud computing falls short of satisfying latency demands due to the transmission of data over long distances and network overload. Bridging data sources and CDCs, it sets up a network of gateways, routers, switches, and compute resources~\cite{nayeri2021application}. The use of fog computing enhances the capabilities of cloud computing due to its minimal latency and cost-effectiveness, as well as the decrease in bandwidth necessary for the transit of data. It is more secure to process confidential information locally at the fog nodes, and if/when needed, only submit trained models---not raw data---to intermediate nodes and eventually the cloud for aggregation, e.g., via federated learning~\cite{patros2023rural}. These applications collect data from various IoT devices to deliver useful insights and deal with latency issues~\cite{mahmud2020application}.
\item \textit{P2P Network}: This network is formed in its most basic form when two or more PCs are linked to one another and exchange resources without passing through a third computer that acts as a server~\cite{suryono2019peer}. A P2P network might be a spontaneous connection, which would consist of two or more computers linked together using a Universal Serial Bus for the purpose of file sharing. In a fixed infrastructure, P2P networking utilizes copper lines to connect six PCs located in a single workplace~\cite{schollmeier2003protocol}. Alternately, a peer-to-peer network may be an ecosystem that is considerably larger in scale and is characterized by the use of specialized protocols and apps to establish direct links between consumers over the web.

\item \textit{Osmotic Computing}: This model has become pervasive in various settings, from urban planning and healthcare to linked vehicles and Industry 4.0~\cite{neha2022systematic}. It lays the groundwork for a system in which vehicles, pedestrians, and urban infrastructure interact and share real-time information to improve traffic flow. As more people use IoT applications housed in different types of networks (cloud, edge, and IoT), it is now clear that the providers who make up the IoT's service ecosystem (data, service, network, and equipment) are all interconnected~\cite{gushev2020dew}. In this setting, buyers and sellers implicitly expect their data and services to be secure and trustworthy. There is no requirement for familiarity with the federated ecosystem (service, data, and network) for users of the IoT apps to connect with many applications using a web-based user interface~\cite{ruggeri2022innovative}. Users send their information to application providers without realizing that those trusted suppliers may share that information with any third parties (such as a company that hosts analytics on the cloud or a company that provides the infrastructure for mobile devices). Security issues may arise for software due to the wide variety of computing devices available from different manufacturers and their presence in an untrusted realm with no overarching authority~\cite{gill2018secure}.
\item \textit{Dew Computing}: It stands out because of its near-complete independence from Internet access, its users' physical closeness to servers, its low latency, outstanding speed, excellent user interface, and adaptability in terms of control available to users~\cite{ahammad2021review}. Instead of serving as a replacement for cloud computing, dew computing serves as a useful supplement. In the not-too-distant future, people throughout the globe might be able to limit their time spent online, increasing their efficiency and effectiveness. Countries have adopted measures to handle the influx of Internet users caused by the COVID-19 blackout. To lighten the Internet's burden, video streaming services are reducing visual quality, while others just update their software outside of peak viewing times. The dew computer's proximity to the user in the design means it can facilitate all electronic interactions with fewer steps and more efficient data transfer~\cite{ahammad2021review}.
\item \textit{Edge Computing}: Since its origins in content delivery networks, distributed computing has matured into the mainstream as an edge computing paradigm that places resources near the client's end. Big data is typically best stored in the cloud, whereas immediate information created by consumers and exclusively for the customer needs computing power and storage on the edge~\cite{shi2016edge}. To accommodate growing mobile user needs, cloud providers have realized they must shift crucial processing to the device. 

With its high performance and low cost, edge computing is a key driver for AI. This can be the most helpful method to see how AI relates to edge computing. Due to the data- and compute-intensive characteristics of AI, edge computing aids AI-powered applications in resolving their technical problems. AI/ML systems consume large amounts of data to discover trends and provide trustworthy recommendations~\cite{mao2017survey}. AI use cases that need video analysis face latency challenges and rising expenses due to the cloud-based transmission of high-definition video data. 

The delay and reliance on central processing in cloud computing are problematic when ML inputs, outputs, and (re-)training data must be handled in real-time. It is possible to perform computation and decisions at the edge, eliminating the need for costly backbone connections and allowing immediate action on the data. Client information regarding location is stored at the edge instead of in the cloud for security reasons. When data is streamed to the cloud, all relevant data and datasets are uploaded. Edge networks for computing have introduced several difficulties associated with infrastructure management because of their dispersed and intricate nature~\cite{luo2021resource}. Managing resources efficiently requires carrying out several tasks. Examples include VM consolidation, resource optimization, energy efficiency, workload prediction, and scheduling. Resource management has historically relied on static, established guidelines, mostly based on operations research methodologies, even in dynamic, rapidly changing settings and in immediate situations. To deal with these issues, especially when choices must be made, AI-based solutions are being used more and more frequently. AI/ML methods have become increasingly common in the past few years~\cite{cao2020overview}. However, selecting where on edge to carry out a task can be challenging, as it requires considering tradeoffs like the volume of data on edge servers and the ability to move users~\cite{kotsehub2022flox}. The cache has to anticipate the consumer's next destination for it to build on the notion of mobility~\cite{almurshed2022adaptive}. It is situated at a suitable edge to cut costs and energy consumption. Several different methods, including genetic algorithms, neural network models, and reinforcement learning, are utilized in this process.
\end{enumerate*}
\begin{itemize}
\item \textit{Mobile Edge Computing}: Mobile Edge Computing---now Multi-access Edge Computing (MEC)---expands its possibilities by introducing cloud computing to the web's edge. Initially targeted solely on the edge nodes of mobile networks, MEC has since expanded its scope to include conventional networks and, ultimately, integrated networks. While typical cloud computing occurs on servers located far from the end-user and devices, MEC enables activities to be carried out at base stations, centralized controllers, and various other aggregating sites on the Internet~\cite {abbas2017mobile}. MEC improves consumer QoE by redistributing cloud computing workloads to customers' individual, on-premises servers, thus relieving congestion on mobile networks and lowering latency~\cite{Du2023}. Innovative applications, services, and user experiences are being unlocked at a dizzying rate thanks to advances in edge data generation, collection, and analysis and in the transmission of data between devices and the cloud~\cite{mach2017mobile}. Because of this, MEC is accessible to consumers and businesses in a wide range of contexts and industries. Integrating MEC into a camera network improves the speed with which data may be stored and processed. With sufficient processing power and bandwidth, data may be immediately analyzed locally instead of being sent to a remote data center~\cite{siriwardhana2021survey}. Self-driving automobiles and autonomous mobile robots (AMRs) are two examples of emerging technologies that require powerful ML to arrive at judgments rapidly. If such decisions take place in a remote data center, only seconds might be the distinction between nearly escaping failures and causing a tragedy~\cite{mao2017survey}. Because the vehicle must avoid hitting pedestrians, animals, and other vehicles, judgments must be made on the vehicle. Machine-to-machine (M2M) communication will be essential to the success of 6G as the forthcoming generation of a global wireless standard and the technological advances that will emerge from it~\cite{akyildiz20206g}.
\end{itemize}

\subsubsubsection{Technologies/Impact Areas} 
The key technologies and affected domains for decentralized computing include:

\subsubsubsubsection{Distributed Ledger Technology}
The computing paradigms of fog, edge, and cloud are currently experiencing explosive growth in both the business and academic worlds. Security, confidentiality, and data integrity in these systems have become increasingly important as their practical applications have expanded~\cite{golec2023blockfaas}. Data loss, theft, and corruption from malicious software like ransomware, trojans, and viruses raise serious considerations in this area. For the system's and most importantly, end-users' sake, it is crucial that data integrity be maintained, and that no data be delivered from an unauthenticated source. Medical care, innovative cities, transport, and monitoring are all examples of applications of critical importance where the margin for error is near zero~\cite{gill2022ai}. 
\begin{itemize}
\item \textit{Blockchain}: Because the majority of edge devices have limited computing and storage capacity, developing an appropriate system for data security, and preserving integrity is challenging. The IoT and other real-time systems have used blockchain technology for data security~\cite{doyle2022blockchainbus}. To store and monitor the worth of an asset over time, a blockchain is, in theory, a set of distributed ledgers. When new information is added to the system, it becomes a block with a Proof of Work (PoW). A PoW is a hash value that cannot be made without changing the PoW of the blocks that came before it in the ledger. Miners create and verify these PoWs while also mining blocks in the Fog network~\cite{zheng2018blockchain}. 

After a miner has completed the PoW, it broadcasts the newly created block into the network, where the other nodes check its legitimacy before joining it in the chain. Also, the fraudulent change of data in a blockchain will not work unless at least half of the copies of the data in question are changed individually by carrying out the same actions. With such a strict time constraint, modifying any data in the blockchain will be extremely difficult. Network nodes must offer route selection, preservation, financial services, and mining for the blockchain to function. Considering these challenges, numerous groups have worked to develop solid frameworks for combining blockchain and fog computing~\cite{al2019blockchain}. 

The majority of these systems employ a dynamic allocation mining technique in which the least-used nodes mine and validate the chains. In contrast, the remaining nodes are employed for load balancing, computation, and data collection~\cite{yang2019integrated, gai2020blockchain}. The blockchain on a large portion of the network is replicated at those nodes if a worker detects an issue in relation to blockchain manipulation or signature forging. Furthermore, blockchains offer public-key encryption with adaptive key exchange for further security. Blockchain is a deceptively simple central notion, but incorporating it into fog computing systems presents several challenges. Cost and upkeep are major factors surrounding storage capacity and scalability. Only complete nodes (nodes that can fully validate the transactions or blocks) store the whole chain, which still results in massive storage needs. Data anonymity and privacy issues are another blockchain shortcoming. Privacy is, therefore, not incorporated into the blockchain architecture; consequently, third-party tools are necessary for accomplishing these crucial requirements~\cite{moqurrab2022deep}. This might result in less efficient applications that demand more resources (both computationally and in terms of storage space) to run. There are still numerous unresolved issues and potential future developments for blockchains in IoT architectures~\cite{gill2019transformative}. 

Insufficient resources are the main barrier to excellent data protection and dependability. Because of resource limits, more complex encryption or key generation cannot be incorporated with these chains of data~\cite{golec2022aiblock}. Only restricted encryption algorithms may be implemented. By considering resource limitations, more effective algorithms may be created. In high fault-rate scenarios, wherein the edge nodes are susceptible to attack at any time, modifying such chains is another essential approach~\cite{kumar2023blockchain}. Network and I/O bandwidth needs are greatly increased due to the necessity of revalidating blocks and copying chains from the primary network. The majority of frameworks additionally use a master-slave architecture, which introduces a potential weak spot. In diverse settings, this is to be expected. The balance between cost and reliability must be meticulously evaluated when considering redundancy~\cite{sharma2022applications}. The blockchain flaws also continue to impact fog architectures. There is a need to develop efficient consensus techniques that can validate blocks with little block sharing and copying. Those curious might learn more about blockchain by reading an in-depth report on the topic.
\end{itemize}

\subsubsubsubsection{Federated Learning}
Data is needed for ML model training, testing, and validation. Information is stored in locations accessible by thousands or millions of users (devices). Rather than sharing the entire dataset required to train a model, federated devices only communicate the parameters specific to that device's instance of the model. The parameter sharing mechanism is defined by the federated learning topology~\cite{li2020review}. Each participant in a centralized topology contributes the parameters of the model to a centralized server, which then trains the centralized model and returns the trained parameters to each participant. Parameters are typically shared among a smaller group of peers in other configurations, including peer-to-peer or hierarchical ones. ML methods that require large or geographically dispersed data sets may benefit from federated learning. However, there is no universally applicable machine-learning solution~\cite{yang2022federated}. Several unanswered questions remain about federated learning that researchers and developers are hard at work trying to address~\cite{zhang2021survey,Jiang_2024}. There are a lot of opportunities for efficient communication in federated learning. This means the master server or other entities acquiring the parameters must be able to cope with occasional interruptions or delays in transmission. Getting all the federated devices to talk to each other and stay in sync is still an open issue~\cite{zhang2021survey}. There is typically a lack of transparency between federated parties and a central server regarding the computing capacity of the federated parties. However, it is still challenging to ensure that the training activities will operate on a diverse mix of devices~\cite{li2020review}. Federated parties' data sets might be quite varied in terms of data amount, reliability, and variety~\cite{kairouz2021advances}. It is challenging to predict how statistically diverse the training data sets will be and how to protect against any detrimental effects this diversity may have. Efficient deployment of privacy-enhancing solutions is required to prevent data loss due to shared model parameters.

\subsubsubsubsection{Bitcoin Currency}
Transaction settlement using blockchain technology was initially proposed with the digital (crypto-)currency Bitcoin. The blockchain is a distributed ledger that verifies monetary transactions using PoW and may be configured to record anything of worth. Blockchains, including bitcoins and cryptocurrencies, are innovative in operating apps across networks~\cite{wu2024privacy}. Designers create smart contracts for Bitcoin money exchanges, which are subsequently carried out on blockchain VMs~\cite{ferdous2021survey}.

Blockchain relies on a decentralized, concurrency-agnostic runtime environment and consensus mechanism. Blocks of data may be disseminated across Bitcoin ledgers via a peer-to-peer network with no requirement for a centralized authority, thanks to the Bitcoin enabling network~\cite{ferdous2021survey}. The data in the blockchain is certified by the members to keep it safe and open, and anybody is welcome to join the network. Cloud computing may use this property, and the security of cloud storage, in particular, can benefit from it. Cloud computing infrastructures enable the execution of complex applications and the handling of massive data sets. Centralized data centers coupled with Fog or IoT devices at the network edge cannot efficiently handle the enormous data storage required to deliver high-availability, real-time, low-latency services~\cite{manimuthu2019literature}. 

A distributed cloud design is required to deal with these problems instead of the more conventional network architecture. Blockchain technology, a fundamental element of distributed cloud systems, offers detailed control over resources by enabling their management through distributed apps~\cite{xu2023survey}. It also allows for the tracking of resource usage, providing both customers and service providers with the means to verify that the agreed-upon QoS is being met. A marketplace is a platform where everyone may promote their computer resources while discovering what they require using AI-based techniques or models of prediction~\cite{wang2023blockchain}. Blockchains, compared to cloud computing, offer fewer computer resources available to run distributed applications, such as less space for storing data, less powerful VMs, and a more unstable protocol. As a result, apps that are sensitive to delay and those that use a lot of resources need to find solutions to these problems~\cite{rahardja2021good}. 

Combining blockchain and cloud computing to develop a blockchain-based distributed cloud can provide novel advantages and solve current restrictions. Data moves closer to its owner and user through Blockchain's distributed cloud, providing on-demand resources, security, and cost-effective access to infrastructure~\cite{golec2021ifaasbus}. In the meantime, the high price and substantial consumption of electricity from clouds may be solved with a blockchain-based distributed cloud. Cloud storage security is another area where blockchain may play a role in the future~\cite{Qu2022ChainFL}. By dividing user data into smaller pieces before storing it everywhere, it is possible to encrypt it further. A small portion of the data is accessible to the hacker, not the entire file. In addition to eradicating data-altering hackers from the network, a backup copy of the data may be used to restore any changes~\cite{wang2023blockchain}. The use of quantum computers to circumvent the mathematical impossibility of modern encryption is one of their most publicized uses. In the meantime, many online publications have predicted the end of Bitcoin and other cryptocurrency use after Google stated it had achieved quantum supremacy.

%\end{enumerate*}

\subsubsubsection{Trends/Observations} 
The main trends and observations regarding decentralized computing are as follows:

 \textit{Serverless Edge Computing}: Serverless' `scale-to-zero' feature, which releases unoccupied containers from the system, works well for energy-conscious IoT scenarios with load-inconsistent applications. On the other hand, fine-grained scaling (i.e., at the function stage) is capable of handling extremely distinct needs and execution settings at the edge~\cite{aslanpour2021serverless}. Many IoT applications rely on instances initiated by sensing or actuating, just like functions in serverless~\cite{ghafouri2022mobile}. However, unlike serverless functions, IoT devices often sense or act only on rare occasions, whereas they sleep the majority of the time to conserve power. So, first, serverless appears to be an ideal paradigm of execution. However, combining serverless, edge computing, and IoT applications is challenging because serverless was originally designed for cloud environments, which do not have the same constraints as edge computing devices \cite{golec2023healthfaas}. In light of this opportunity, it is essential to combine serverless, edge computing, and IoT applications to address this challenge. This is crucial to be addressed, as the fact is that although this adaptation looks needed and helpful, its practicality necessitates comprehensive inspections to avoid ramifications.
%\end{itemize}

\subsubsection{Hybrid Computing}
It involves combining both a centralized network and a decentralized network. In this section, we discuss the main focus or paradigms, technologies or impact areas, and various trends or observations within hybrid computing.

\subsubsubsection{Focus/Paradigms} 
The following are the main focus or paradigms for hybrid computing:

\textit{Fog-Cloud-Edge Orchestration}: Increasingly, IoT technologies are required in daily life. Smart cities, automated manufacturing, virtual reality, and autonomous cars are just a few instances of the vast variety of sectors where the application of these technologies has been rising quickly~\cite{svorobej2020orchestration}. This type of IoT application frequently necessitates access to heterogeneous distant, local, and multi-cloud compute resources, in addition to a globally dispersed array of sensors. The expanded Fog-Cloud-Edge orchestration paradigm is born from this. This new paradigm has made it a necessity to expand application-orchestration needs (i.e., self-service deployment and run-time administration) beyond the confines of a purely cloud-based infrastructure and across the full breadth of cloud or edge resources. Recent years have seen an increased focus on the research and advancement of orchestrating platforms in both business and academic settings as a means of meeting this need.
%\end{itemize}

\subsubsubsection{Technologies/Impact Areas} 
The key technologies and affected domains for hybrid computing include:

\begin{enumerate*}[label =\color{black} \it \arabic*),itemjoin=\\\hspace*{\parindent}] 
\item \textit{Cryptocurrencies}: Decentralized networks with powerful computational power were pioneered by cryptocurrencies. There is no centralized authority that controls the cryptocurrency market or issues new cryptocurrencies. Bitcoin, the first decentralized digital currency, was launched in 2009 and employs blockchain technology to record transactions and save user histories~\cite{hardle2020understanding}. Blockchain Explorer and similar tools reveal Bitcoin's decentralized network activity as it moves from one wallet to another, and they also reveal the activity of other cryptocurrency networks. There is no equivalent technology that would enable such transparency in the private banking business, nor would such a publication ever be made public. Decentralization design incorporates many additional features that make it hard for bad actors to forge bitcoin or steal from user accounts, such as synchronizing the blockchain across all machines on the network~\cite{weichbroth2023security}. Bitcoin and other cryptocurrencies are required to function on decentralized networks: A blockchain does not have a central controlling computer or administrator.
\item \textit{Machine Economy}: The emerging machine economy refers to the exchange of resources (such as power, data storage, processing power, currency, and network connections) in the upcoming global networks of computers~\cite{schweizer2020extent}. Together, the data centers that power the cloud, the web, and monetary exchanges, form a network that will support the machines that power the future economy. This is the time when AI willfully conceals or exaggerates its powers. AI conceals and safeguards limited supplies to protect the crucial scarce resource of computation cycles used to generate AI insights. The organization is guarding the computation cycles used to generate AI insights, which are the most crucial scarce resource in this case. Lies, trickery, and barter to coax AI into parting with its limited resources will become an increasingly hot issue in the coming years~\cite{khan2022review}. To prevent itself from being overused, AI will have to resort to dishonest behavior. The machine economy is going to be among the most significant developments to come for human culture; and will be among the hottest topics of the emerging payment and AI technologies needed to fund future interstellar and interplanetary travel.
\end{enumerate*}

\subsubsubsection{Trends/Observations} 
The main trends and observations regarding hybrid computing are as follows:

\textit{Distributed Computing Continuum}: Emerging from the convergence of IoT, edge, fog, and cloud computing, Distributed Computing Continuum Systems (DCCS) represent a novel computing paradigm that harnesses the collective power and heterogeneity of these diverse computing tiers to address the demanding computational requirements of future applications~\cite{dustdar2022distributed}. These applications, ranging from autonomous vehicles and e-Health to smart cities, holographic communications, and virtual reality, demand unprecedented levels of computational power, low latency, and efficient data management. Achieving these stringent requirements necessitates seamless integration and collaborative operation among all computing tiers, transforming the underlying infrastructure into a unified, intelligent system.
As exemplified by edge and fog computing, the underlying infrastructure of DCCS plays a pivotal role in determining its performance. This geographically distributed, heterogeneous, and resource-constrained infrastructure poses significant challenges, needing new approaches that can dynamically adapt to application and user demands~\cite{casamayor2023fundamental}. Cloud-centric methodologies, often tailored to cloud-specific assumptions, fall short in addressing the characteristics of edge, fog, and DCCS environments.

To address these challenges, DCCS advocates for decentralized intelligence, empowering each component of the underlying infrastructure to make autonomous decisions based on its specific tasks and local conditions~\cite{donta2023exploring}. This approach leverages the concept of service level objectives (SLOs), well-established in cloud computing, to define the operational goals of each component of the system. By modularizing and distributing SLOs across the system, a DCCS can achieve scalable intelligence within its infrastructure.
Further, incorporating the Markov Blanket concept into SLO management enables causal filtering, ensuring that only conditionally dependent variables are considered when making decisions. This selective filtering, coupled with causal inference or active inference, empowers each component to make informed decisions independently, adapting to its dynamic environment and the overall system's requirements~\cite{morichetta_roadmap_2021}. This loosely-coupled architecture fosters a resilient and adaptive DCCS, capable of catering to the diverse and evolving demands of future applications.

%\end{itemize}

\subsection{Computational Methodologies: Parallel vs. Sequential Computing}
Parallel computing implies a computer model wherein numerous tasks are completed concurrently, employing a number of processors or threads~\cite{beasley1998new}. In this paradigm, many processes run concurrently and their outputs are pooled. Tasks can be conducted in parallel instead of sequentially, potentially reducing execution times.

\subsubsection{Parallel Computing}
In this section, we discuss the main focus or paradigms, technologies or impact areas, and various trends or observations within parallel computing. 

\subsubsubsection{Focus/Paradigms} 
The following are the main focus or paradigms for parallel computing. 

\textit{Simultaneous Data Processing}: In order to handle many parts of a task at once, parallel processing employs multiple processors, or CPUs. By breaking down large computations into smaller ones, systems may drastically speed up their execution~\cite{beasley1998new}. Parallel processing is possible on current computers with multiple cores and on any machine with more than one CPU. Multi-core processors are embedded processors containing two or more CPUs for increased performance, lowered energy use, and more efficient handling of many tasks. Two to four cores are common in modern computers, with some models supporting up to 12. Modern computers commonly use parallel processing to complete complex processes and calculations. At the most basic level, sequential and parallel-serial processes differ in how registers are employed. Shift registers work in series, computing every bit one at a time, while registers with concurrent loading handle each bit of a word concurrently~\cite{aminizadeh2023applications}. Using multiple functional units that can execute identical or distinct tasks in parallel enables the management of more complex parallel processing.
%\end{itemize}

\subsubsubsection{Technologies/Impact Areas}
The key technologies and affected domains for parallel computing include:

\begin{enumerate*}[label =\color{black} \it \arabic*),itemjoin=\\\hspace*{\parindent}] 
\item \textit{ASICs}: Application-Specific Integrated Circuits (ASICs) are integrated circuits designed for specific uses. As their name suggests, ASICs are limited to a single function. They provide a single function and are consistent throughout their service life~\cite{ferraz2021survey}. ASICs are semiconductor devices and circuitry developed to carry out a particular task. In contrast to mainstream processors, including CPUs and GPUs, both the speed and the energy efficiency of ASICs are optimized to fit the needs of a specific application~\cite{petrou2022first}. Their excellent performance, minimal energy use, and small form factor make them ideal for mass-produced goods that can afford the higher bespoke design costs.
\item \textit{FPGA}: A Field Programmable Gate Array (FPGA) is a semiconductor that can be programmed to provide unique logic for use in both early system prototype design and the last version of a system to circumvent obsolescence~\cite{ferraz2021survey}. In contrast to other bespoke or semi-custom integrated circuits, FPGAs can be easily reprogrammed by a software update to meet the changing requirements of the larger system they are integrated into, using hardware design languages, such Verilog and \textit{Very High-Speed Integrated Circuit Hardware Description Language (VHDL)}~\cite{murray2020vtr}. Nowadays, most rapidly expanding applications are perfect fits for FPGAs, which include edge computing, AI, network security, 5G, industrial control, and automated machinery.
\end{enumerate*}

\subsubsubsection{Trends/Observations} 
The main trends and observations regarding parallel computing are as follows:

\begin{enumerate*}[label =\color{black} \it \arabic*),itemjoin=\\\hspace*{\parindent}] 
\item \textit{Neuro-symbolic AI}: Advances in deep learning techniques have unlocked a few of AI's enormous possibilities. Consequently, it is now obvious that these methods are at a breaking point and that such sub-symbolic or neuro-inspired solutions only function effectively for particular kinds of issues and are typically opaque to both analysis and comprehension~\cite{hitzler2022neuro}. However, symbolic AI methods, founded on rules, logic, and reasoning, perform significantly better in terms of openness, comprehensibility, authenticity, and reliability than sub-symbolic methods. A new path termed neuro-symbolic AI was recently recommended, integrating the effectiveness of sub-symbolic AI alongside the visibility of symbolic AI~\cite{gaur2022knowledge}. This synergy has the potential to yield a new generation of AI devices and platforms that are both comprehensible and expansion-intolerant and can combine logic with learning in a generic fashion.
\item \textit{Scalability}: The most important advantage of scalable design is improved efficiency, as well as the capacity to deal with sudden spikes in traffic or severe loads with little to no warning ~\cite{du2023computation}. An application or online company may continue to operate smoothly during busy periods with the assistance of a scalable system, preventing businesses from incurring financial losses or suffering reputational harm~\cite{xu2022coscal}. If a system is organized into component services (for example, using the microservices system design), monitoring, updating features, troubleshooting, and scaling may become simpler tasks.
\end{enumerate*}

\subsubsection{Sequential Computing}
In this section, we discuss the main focus or paradigms, technologies or impact areas, and various trends or observations within sequential computing.

\subsubsubsection{Focus/Paradigms} 
The following are the main focus or paradigms for sequential computing:

\textit{One-process-at-a-time execution}: In the context of computing, sequential computing describes a paradigm in which operations are carried out in a certain order, with the output of one operation feeding into the data being the input of the subsequent one~\cite{cuadrado2000intelligent}. A single processor carries out all of the model's tasks in the sequence specified by the code.
%\end{itemize}

\subsubsubsection{Technologies/Impact Areas} 
The key technologies and affected domains for sequential computing include: 
 
\textit{Traditional Von Neumann Architecture}: This architecture is a sequential computing-based concept for digital machines. This system includes a CPU, RAM, and I/O devices, all interconnected by a bus~\cite{zhang2013transparent}. The CPU of a system based on the Von Neumann architecture processes instructions sequentially, feeding the output of one into the input channel of the subsequent one~\cite{kimovski2023beyond}.
%\end{itemize}

\subsubsubsection{Trends/Observations} 
The main trends and observations regarding sequential computing are as follows:

\begin{enumerate*}[label =\color{black} \it \arabic*),itemjoin=\\\hspace*{\parindent}] 
\item \textit{In-Memory Computing}: In-memory computing is a method used to perform computations solely in memory (like RAM). This word usually refers to massive and complicated computations that must be executed on a cluster of computers using specialized systems software~\cite{cuadrado2000intelligent}. As a clustering system, the machines pool their RAM, so the computation is effectively done across machines and uses the combined RAM capacity of all the machines collectively.

\item \textit{Energy-efficiency}: Power effectiveness and sustainability have emerged as major issues for HPC systems as their processing capacity increases~\cite{jiang2019adaptive}. To reduce electrical usage while increasing computational performance, scientists are inventing environmentally friendly hardware layouts, investigating innovative cooling strategies, and fine-tuning algorithms. The general efficiency of HPC systems is being improved by the development of energy-aware scheduling and utilization strategies.

\item \textit{Performance Optimization}: Since single-processor efficiency can no longer develop at a rapid pace, the era of the single-microprocessor computer is coming to an end. It's time for a new era in computing when parallelism takes centre stage and sequential computing takes a back seat~\cite{bufistov2007general}. There are still significant scientific and engineering obstacles to overcome, but now is a good moment to try new approaches to computer programming and hardware design. Various computer architectures have emerged, each tailored to certain performance and efficiency goals. The next wave of discoveries will certainly necessitate enhancements to computer hardware and software~\cite{aslanpour2020performance}. No one can say for sure if we'll succeed in making parallel computing as mainstream and user-friendly as yesterday's peak sequential single-processor computer systems in the field of computing. Innovative novel applications that motivate the computer business will slow down if parallel programming and associated software activities don't become popular, and if creativity slows down across the economy as a whole, many other sectors will suffer as well~\cite{zhang2018performance}.
\end{enumerate*}

\subsection{Computing Trends and Emerging Technologies}
New computing trends and emerging technologies continue to advance the field of computing, improving the adaptability, self-management, and sustainability of many types of industrial systems.

\subsubsection{Advanced Computing Styles and Trends}
In this section, we discuss advanced computing styles and trends and their related technologies and paradigms. 

\subsubsubsection{Focus/Paradigms}
The following are the main focus or paradigms for advanced computing styles:

\textit{Quantum AI}: Quantum computing is attractive because it is a unique innovation that can radically change AI and computing in general. In this section, we look into what quantum computing can do and how it can affect AI and the wider economy. The implications of this computing method might have far-reaching effects on several facets of our cultural and financial lives~\cite{gill2022ai}. The widespread impact of AI suggests that combining it with quantum computing might unleash dramatic change in the field of AI~\cite{gill2021quantum}. 

Several algorithms that made it possible to do tasks previously thought impossible for conventional computers emerged in the wake of the foundational studies that formalized the notion of a quantum computer~\cite{singh2021quantum}. The development of Shor's algorithm, an effective method for dividing enormous amounts of data, has bolstered research into quantum computing and quantum cryptography. Yet, existing cutting-edge technologies are not yet accurate enough to execute Shor's algorithm successfully, which requires a degree of precision for performing register initialization, quantum operations on multiple qubits, and storing quantum states. It is also crucial to remember that quantum computers have particular limits~\cite{singh2022quantum}. The acceleration afforded by quantum computers grows exponentially compared to the amount of time a conventional computer takes (Grover's method); hence, it is not predicted that it will effectively solve NP-hard efficiency issues. The benefits of quantum computing, such as quantum superposition and entanglement, typically vanish rapidly with the complexity and magnitude (i.e., the number of quantum systems involved) of the underlying hardware, making the process of designing a quantum computer non-trivial. Despite this, the curiosity of significant technologically advanced players (IBM, Microsoft, Google, Amazon, Intel, and Honeywell) has skyrocketed in the past few years, and a plethora of fresh startups have emerged to propose remedies for quantum computing using technologies as diverse as superconducting devices, encased ions, and integrated light circuits. Corporations like these are among the numerous that are investing in quantum research and development at the moment~\cite{de2021materials}. 

Although there are many obstacles to overcome, the Google AI team has achieved considerable strides in the past few years, gaining a quantum edge by developing Sycamore, a programmable quantum computer. Similarly, IBM has now launched the Eagle chip, the first quantum computer with more than 100 qubits of hardware~\cite{smith2022scaling}. This is only the beginning of an intensive research and development program, with the tech giant hoping to increase the number of qubits to over a thousand by 2024~\cite{gill2022quantum}. But as was previously stated, protecting these devices from ambient noise is a significant constraint when trying to retain the subtle characteristics of composite quantum states while still allowing for coherence in quantum development. Because of this, a quantum computer's components require ultra-low temperatures in the order of fractions of a Kelvin, which presents hurdles for both device design and material development~\cite{spivey2021high}.

%\end{itemize}

\subsubsubsection{Trends/Technologies} 
The main trends and technologies regarding advanced computing styles are as follows: 

\begin{enumerate*}[label =\color{black} \it \arabic*),itemjoin=\\\hspace*{\parindent}] 
\item \textit{Edge AI}: Recent advancements in AI efficiency, the rise of IoT devices, and the emergence of edge computing have all unleashed the promise of edge AI. This has opened up previously unimaginable uses for edge AI, such as helping radiologists make diagnoses, assisting in driving cars and fertilizing crops~\cite{singh2023edge}. Since its inception in the mid-1990s---paired with the emergence of content delivery networks that utilize edge servers positioned near users to stream online and gaming video---edge computing has been the subject of much discussion and adoption by professionals and businesses. Almost every sector today has tasks that may benefit from adopting edge AI. 

In truth, edge applications are driving the next generation of AI computing, which will improve people's lives in various settings, such as at home, at work, at school, and on the road. AI at the edge refers to the application of AI to physical devices. In contrast to storing all of an organization's data in a single centralized spot, such as a cloud provider's data centre or a private data warehouse, ``Edge AI'' allows for AI calculations to be performed close to the users at the network's edge. Because the Internet is accessible all across the globe, any area might be thought of as its outskirts. Omnipresent traffic signals, autonomous equipment, and mobile phones are just a few examples. It might also be anything from a shop to a factory to a healthcare facility. Companies of all sizes strive to automate more of their processes because doing so improves productivity, effectiveness, and safety~\cite{nandhakumar2023edgeaisim}. Computer software may aid with this through the ability to recognize patterns and dependably carry out identical tasks repeatedly. However, it is challenging to fully convey them in a system of algorithms and regulations because the world is unpredictable and human actions cover infinite circumstances. Today, as edge AI has progressed, robots and devices can work with the ``intelligence'' of human cognition no matter what they are. Intelligent IoT apps driven by AI may learn to adjust to novel circumstances and effectively complete identical or similar tasks~\cite{Xue2022}. Substantial progress in important areas has allowed for the practical deployment of AI models at the edge. 

Furthermore, developments in neural networks, along with other areas of AI, have laid the groundwork for universal ML~\cite{lee2018techology}. Many companies are finding that they can successfully train AI models and put them into action at the edge. AI in the periphery requires widely distributed computing resources. Recent advancements in enormously parallel GPUs are currently used to run neural networks. The development of devices connected to the IoT is partly responsible for the present age's unparalleled surge in data volume~\cite{ding2022roadmap}. The development of sensors, smart cameras, robots, and other data-gathering equipment has made it possible to begin using AI models at the edge in nearly all facets of business. The increased speed, dependability, and security that 5G/6G is delivering to the battleground are also helping IoT use cases~\cite{shen2021holistic}.

\item \textit{Biologically-inspired Computing}: The term ``bio-inspired computing'' refers to creating computer systems by drawing inspiration from the natural world. As an aside, computer science is also used to model and understand biological processes~\cite{zangeneh2021analogue}. 

Computing architectures that take cues from nature can function as autonomous, flexible networks. Similarly, bio-inspired computing offers a fresh perspective on AI by building modular, self-improving systems~\cite{murugesan2023comparison}. Swarm intelligence refers to the ability of swarms of autonomous entities to generate intelligence by collaborating in ways reminiscent of the behavior of bees or ants. Biologists, software engineers, computer scientists, physicists, mathematicians, and geneticists all work together on the subject of bio-inspired computing~\cite{kar2016bio}. Compared to their digital counterparts, biological systems have several distinct benefits. AI has advanced thanks to incorporating many concepts originally derived from natural processes into machine learning. Adaptable and responsive autonomous robots might be extremely useful in high-risk settings like conflict zones and hazardous clean-up activities~\cite{zhang2020neuro}. 

Tasks like crop pollination might be performed by swarms of small robots. Bio-inspired technology is being used in cognitive modelling by developing artificial neural network systems based on neuron function within the brain. Training, growing, and collaborating on computer chips is becoming a reality~\cite{xu2022esdnn}. When these nodes are linked by self-organizing wireless links, they form a system well adapted to modelling issues with several basic causes~\cite{kar2016bio}. Self-learning and reconfigurable chips mean less time spent loading software and more time spent getting things done. Such systems might help explain the propagation of ideas through a community or construct a model of brain function that reflects true biological processes. 

The use of DNA in natural computing is a topic of current study. Data storage, covert messaging, and even computation are all possibilities that have been proposed by DNA bioinformatics studies DNA~\cite{denkena2021reprint}. DNA molecules may also form practical structures by self-assembly. The computer hardware, such as switches, CPUs, and timers, might be replaced by biological components. It is already possible to employ some biological substances in electronics. Even internal cell programming for purposes like medication secretion is feasible.

\item \textit{Explainable Artificial Intelligence (XAI)}: Successful completion of computer engineering tasks depends on wise decision-making. Can workloads be reliably executed on an automated system? Is there any way to understand how the trained models came to their conclusions? Problems like this are typical and must be solved until any computer can be used in action~\cite{gill2022ai}. Incorrect decision-making about such complicated and cutting-edge technology is costly in terms of resources and money. Many AI/ML implementations in computer systems have improved resource utilization and energy usage through better decision-making. However, the forecasts made by these AI/ML models for computing devices are still not usable, interpretable, or implementable. Such restrictions are a common problem for AI/ML models~\cite{dwivedi2023explainable}. Most current research has focused on clarifying how QoS is accomplished, even though QoS remains a top priority. Is there anything academics can do to help the IT industry move forward? Therefore, when attempting to make educated judgments on handling resources (a prime manifestation of AI for computing), a solid grounding in Explainable AI (XAI) and experience with XAI methods and tools is required~\cite{tosun2020histomapr}. Forecasting of resource and power consumption and SLA variances, as well as the implementation of promptly proactive action to resolve these concerns, are examples of the types of Explainable AI techniques that may be used. XAI forecasting algorithms must be correctly developed to make computing more practical, explicable, and deployable~\cite{arrieta2020explainable}. 
\item \textit{Semantic Web and Decentralized Systems Integration}: Fog computing has emerged as a software engineering culture and practice that combines at least five different technology types: IoT, AI, Cloud-to-Edge Computing, Blockchain, and Digital Twins~\cite{KOCHOVSKI2019747}. Various recent projects have presented their vision of integration between the Semantic Web and decentralized systems, for example, networks based on Blockchain technologies~\cite{SHKEMBI2023100809}. Here, the main challenge is to achieve a new generation of trustworthy, sustainable, human-centric, performant, and scalable smart applications. 
\item \textit{Quantum Internet}: It is an ecosystem enabling quantum devices to communicate and share data in a setting that uses quantum physics' peculiar rules. In principle, this would grant the quantum Internet hitherto unattainable skills via standard web apps~\cite{kumar2022securing}. Quantum devices, such as a quantum computer or a quantum processor, may generate the quantum states of qubits, which can then be used to encode information. Sending qubits over a network of physically distinct quantum devices is, in essence, what the quantum Internet will be all about. Importantly, this will occur because of the strange characteristics of quantum states. That probably sounds like the conventional web~\cite{corcoles2019challenges}. However, if one wants to transmit qubits around, then they need to use a quantum channel instead of a conventional one, which requires making use of the peculiar behavior of particles at the subatomic level—--the so-called ``quantum states'' that have both fascinated and perplexed scientists for decades. Information transmission on the quantum Internet is based on quantum physics, a completely foreign field~\cite{singh2021quantum}. One may need to set aside all knowledge of classical computing to comprehend the quantum ecology of a possible Internet 2.0. One could imagine that their favorite web browser will not have much in common with the quantum Internet~\cite{gill2022ai}.
\end{enumerate*}

\subsubsection{Industry and Sustainability Trends}
In this section, we discuss industry and sustainability trends and their related technologies and paradigms. 

\subsubsubsection{Focus/Paradigms} 
The following are the main focus or paradigms for industry and sustainability trends:

\textit{Carbon-Neutral Computing}: The expansion of the computer age is an important factor in the data centre industry's advancement; however, the push towards carbon neutrality is a more dramatic paradigm change and the industry's biggest challenge to date. Large-scale cloud providers have pledged to attain zero emissions on all initiatives by 2030~\cite{seto2021low}. The fight against climate change must include data centers. Everything from everyday conveniences like Internet banking and shopping to cutting-edge technologies like machine learning, quantum technology, and autonomous vehicles would be impossible without them. There is no denying of the ever-increasing need for data centers. Nevertheless, because of the damage they cause to the natural world, they also attract greater scrutiny~\cite{cao2022toward}. A sustainable future with a zero-carbon footprint is possible because of these advancements in electricity, water effectiveness, and land utilization. Online conferences and handheld gadgets make it feasible for individuals to work from their homes and cut transit carbon emissions; however, each bit of data has a carbon footprint of its own~\cite{senthilkumar2023enhancement}. Therefore, whereas electronic devices provide opportunities to enhance our oversight of water and materials and to support sustainable economic growth, simply sending a message provides for the challenging environmental impact of data. However, this may differ greatly depending on the spot and efficiency of the data centers that deal with traffic~\cite{kurniawan2023decarbonization}. Crucially, as globalization brings online amenities to more societies, physical infrastructure, such as data centers, must grow to accommodate an increase in consumers, a majority of whom will be in regions around the globe that currently lack access to green power availability.
%\end{itemize}

\subsubsubsection{Trends/Technologies} 
The main trends and technologies regarding advanced computing styles are as follows: 

\begin{enumerate*}[label =\color{black} \it \arabic*),itemjoin=\\\hspace*{\parindent}] 
\item \textit{Industry 4.0}: The Fourth Industrial Revolution, or Industry 4.0, reshapes how goods are made, enhanced, and disseminated. Emerging innovations such as the IoT, cloud computing, analytics, and AI/ML are being incorporated into manufacturing facilities and processes~\cite{aceto2019survey}. Advanced sensors, software with embedded capabilities, and robots are used in these ``smart industries'' to gather information for more informed decision-making. When data from manufacturing operations is combined with data from Enterprise Resource Planning (ERP), supply chain, customer service, and other corporate systems, information that was previously kept separate can be seen and understood in completely new ways, which leads to even more value being created~\cite{aceto2020industry}. 

Improved efficiency and responsiveness to clients is made possible by the advent of technological innovations such as enhanced automation, predictive maintenance, and automatic optimization of process enhancements~\cite{teoh2021iot}. To enter the fourth industrial revolution, the manufacturing sector must embrace the development of smart factories. The ability to see industrial assets in real-time and access preventative maintenance tools may be gained by analyzing the massive volumes of big data generated from sensors on the production line. Smart factories implementing cutting-edge IoT technology see gains in output and quality~\cite{zheng2021applications}. 

Manufacturing inaccuracies and costs can be reduced by using AI-powered visual insights instead of traditional business models for human inspection. Quality assurance staff may monitor production operations from almost any location with minimal expenditure using a smartphone linked to the cloud. Companies may save money on costly repairs by identifying problems early on with the help of ML algorithms~\cite{qu2020blockchained}. Any business operating in the industrial sector, from individual to process production and even in the energy and mining industries, may use the ideas and tools of Industry 4.0.

\item \textit{Digital Twins}: A digital twin is a computerized model of and connected with a real-world object that may be used to test and improve its design, performance, and usability~\cite{yu2022energy}. Smart sensors embedded in the object capture data in real-time, allowing a digital depiction of the asset to be produced~\cite{alkhateeb2023real}. The model may be used through an asset's lifespan, from development and testing to actual usage, revamping and eventual retirement. To create a digital representation of a physical object, digital twins utilize many technologies. The term ``IoT'' describes the network of interconnected devices and the underlying infrastructure that enables them to exchange data and instructions with one another and the cloud as a whole. With gratitude to the introduction of affordable computer chips and high-bandwidth connectivity, one can now have trillions of gadgets hooked up to the global web. Digital twins use data from IoT sensors to replicate physical properties in a virtual form~\cite{mihai2022digital}. The information is sent into a system or panel to be viewed as it changes in real time. Studying, solving issues, and pattern recognition are just a few examples of the kinds of cognitive challenges that AI seeks to address~\cite{wang2023survey}. AI/ML-based algorithms and statistical models let machines do tasks with little to no human help. They do this by relying on patterns of observation and inference. Machine learning techniques used in digital twins process enormous amounts of sensor data, allowing for the identification of data patterns. Optimization of performance, servicing, emissions outputs, and efficiency may all be gained using data insights provided by AI/ML~\cite{kor2023investigation}. There are several significant distinctions between digital twins and modelling: even though both leverage virtual model-based simulations, a digital twin maintains a two-way connection and can affect the physical object. Offline optimization and the design process are two common applications of simulation. Developers use simulators to test out different iterations of a product. On the contrary, digital twins are interactive and dynamically updated virtual worlds. Both their scope and their utility have increased.
\end{enumerate*}

\subsubsection{Adaptive and Self-Managing Systems}
In this section, we discuss adaptive and self-managing systems and their related technologies and paradigms. 

\subsubsubsection{Focus/Paradigms} 
The following are the main focus or paradigms for adaptive and self-managing systems:

 \textit{Autonomic Computing}: IBM's autonomic computing program was one of the earliest worldwide efforts to develop computing systems with little human intervention required to accomplish predetermined goals~\cite{kephart2003vision}. It was primarily based on findings about how human nerves and thinking work and how they are coordinated---bioinspiration, as discussed above. In autonomic computing, researchers explore how software-intensive systems can make decisions and act without human interaction to reach the (user-specified) ``administration'' objectives~\cite{singh2015qos}. The concept of control for closed- and open-loop systems has significantly impacted the foundations of autonomic computing~\cite{singh2017star}. Multiple independent control networks may coexist in practice inside complex systems. The integration of ML and AI to enhance resource utilization and efficiency at scale remains an important obstacle regardless of investigations into autonomic frameworks to handle computing resources, from a single resource (e.g., a web server) to resource groupings (e.g., several servers inside a CDC)~\cite{gill2022ai}. Autonomous and self-managing systems can be implemented on a spectrum from fully automated to partially automated with human oversight through the use of AI/ML to improve the efficiency and performance of the computing systems.
%\end{itemize}

\subsubsubsection{Trends/Technologies} 
The main trends and technologies regarding adaptive and self-managing systems are as follows:

\textit{SDN-NFV}: The explosion of IoT devices and the concomitant flood of sensor data enable knowledge-driven IoT applications, including connected cities and smart agriculture~\cite{mekki2022software}. To begin providing such services, one must develop a data-gathering method that is flexible enough to adapt to shifting conditions in the field. Network programmability (SDN or NFV) enables the easy reconfiguration of IoT networks~\cite{poutievski2022jupiter}. Current SDN/NFV-based approaches in the IoT environment nevertheless fail owing to a shortage of knowledge of resources and overhead, as well as incompatibility with conventional protocols~\cite{buyya2018manifesto}. This void must be filled by prioritizing resource and power limitations in the creation of SDN/NFV-enabled IoT nodes and network protocols. Assigning traffic sources to those Virtual Network Functions (VNFs) across the most efficient paths, with sufficient energy and network reliability, may maximize the number of active NFV nodes~\cite{casamayor2023fundamental}.
%\end{itemize}

\textbf{Summary:} Table~\ref{tab:table2} lists a summary of open challenges and future directions in Paradigms/ Technologies/ Impact Areas, along with recommendations for further reading. Table~\ref{tab:table3} lists the summary of Trends/Observations for modern computing along with the recommendations for future reading. 

\begin{table*}[!t]
\caption{Summary of open challenges and future directions in Paradigms/Technologies/Impact Areas along with further reading}
\label{tab:table2} %shows the 
\centering
\begin{tabular}{p{3cm}|p{11.5cm}|p{3cm}}
\bottomrule
\textbf{Paradigms/ Technologies/ Impact Areas} & \textbf{Open Challenges and Future Directions} & \textbf{Further Reading} \\
\hline
Cloud Computing &  What are the tradeoffs that need to be established between the various QoS requirements brought on by the large variety of IoT applications operating on cloud systems?  &  ACM CSUR \cite{buyya2018manifesto} \\
\hline
Autonomic Computing &  What additional problems may be addressed by an autonomic computing expansion that is based on AI/ML as the number of IoT and scientific workloads increases? &  Elsevier IoT \cite{gill2022ai}\\
\hline
Mobile Cloud Computing &  How would AI-based deep learning algorithms be used to anticipate the resource demands beforehand for diverse geographic resources needed for mobile cloud computing, requiring new strategies for provisioning and scheduling resources?  & ACM CSUR \cite{ren2019survey} \\
\hline
Green Cloud Computing &   How can improved methods for effective data encoding for lower bandwidth usage and energy-effective transmission in data-intensive IoT devices make cloud computing more environmentally friendly?  & ACM CSUR \cite{gill2018taxonomy} \\
\hline
Fog Computing &  How can AI approaches be utilized to properly schedule tasks when working in locations with varying amounts of fog resources?  & Elsevier JPDC \cite{singh2021fog} \& IEEE COMST \cite{10335918}\\
\hline
Edge Computing  &  In what ways edge computing can be utilized to boost power and resource utilization, hence enhancing QoS? &  IEEE COMST \cite{luo2021resource}  \cite{10335918} \\
\hline
Mobile Edge Computing  &  How can novel resource provisioning and scheduling policies be developed for mobile edge computing that makes use of AI-based deep learning approaches to forecast the resource requirements beforehand for resources that are located in different locations?  &   IEEE COMST \cite{siriwardhana2021survey} \cite{10335918} \& ACM CSUR \cite{ren2019survey}\\
\hline
Serverless Computing &  How to reduce the cold start time and increase scalability using serverless edge computing? &  IEEE TSC \cite{li2022serverless} \& ACM CSUR \cite{shafiei2022serverless} \\
\hline
Osmotic Computing   &  How can osmotic computing improve resource availability or performance at the network edge while moving services from the data center to the edge for AI/ML-driven adaptive administration of microservices? & ACM TOIT \cite{neha2022systematic}\\
\hline
Dew Computing  & How should dew computing allow a highly scalable method that can increase or reduce the real-time demands of performing operations at runtime via utilizing AI? &  Elsevier IoT \cite{gushev2020dew} \\
\hline

Programming Models &   How to select a programming model that efficiently gathers data when and where it is needed while keeping complexity low relative to the total number of processors at hand? & Procedia Computer Science \cite{jackson2015survey} \\
\hline
Virtualization &   How can unbreakable security for VMs be ensured if consumers do not follow recommended practices when it comes to login credentials, installations, and other operations? & ACM CSUR \cite{alam2020survey} \\
\hline

IoT & How to ensure that an SLA is upheld while responding to customer requests as quickly as possible using IoT applications? & IEEE COMST \cite{stoyanova2020survey} \\
\hline

Integrated Computing  &  How may QoS characteristics change if communication between layers in a fog-edge/cloud computing paradigm is improved? & ACM CSUR \cite{yang2019integrated} \&  Elsevier 
 FGCS \cite{botta2016integration}\\
\hline

Connectivity/ Networking  &   How can satisfying the demand or need for network solutions enabling high performance, resilience, dependability, scalability, adaptability, and cybersecurity remain constant? & ACM CSUR \cite{ren2019survey} \& IEEE COMST \cite{wang2017integration}\\
\hline

Container Technologies &   How can the QoS in data processing be enhanced by leveraging containers with virtualization?  & Springer JoS \cite{bentaleb2022containerization} \& Wiley CCPE \cite{casalicchio2020state} \\
\hline
Microservices  &  How to handle errors, ensure data integrity, and communicate effectively amongst services in a distributed system using a microservice architecture? & IEEE TSC \cite{al2022ai}\\
\hline
Software-defined Networks  &   What are some ways in which SDN might help minimize power usage in cloud and edge computing? & Wiley ETT \cite{mekki2022software}\\
\hline
Distributed Ledger Technology (Blockchain) &  How can distributed ledger technology (Blockchain) be utilized to secure the data for IoT applications?  & IEEE COMST \cite{yang2019integrated} \cite{gai2020blockchain}\\
\hline
Federated Learning  &   How could companies ensure privacy in federated learning services, which differ from learning in data centers in that users' data is disclosed to third parties or the centralized server while exchanging model changes during the training stage? & Elsevier KBS \cite{zhang2021survey} \& CIE \cite{li2020review} \\
\hline
Software Engineering  &  How can fault tolerance be improved in computing systems dynamically without manually writing the software code by utilizing AI to ``automatically'' diagnose and fix an error?  & Elsevier JSS \cite{arvanitou2021software}\\
\hline

Distributed Computing Continuum Systems & How can Distributed Computing Continuum Systems consider all computing tiers as a single system and optimize future applications in a decentralized manner? &  IEEE TKDE \cite{dustdar2022distributed} \\
\hline

\end{tabular}
\end{table*}

\begin{table*}[!t]
\caption{Summary of Trends/Observation for modern computing along with future reading}
\label{tab:table3}
\centering
\begin{tabular}{p{2cm}|p{11.5cm}|p{2.5cm}}
\bottomrule
\textbf{Trends/ Observation} & \textbf{Open Challenges and Future Directions} & \textbf{Further Reading}\\
\hline
AI-driven Computing &  How to optimize the management of resources using the latest AI/ ML models in computing systems? &  Elsevier IoT \cite{gill2022ai}\\
\hline
Large Scale Machine Learning &    How can businesses mitigate the risks associated with the proliferation of sensitive information that arise as a result of the proliferation of data produced by AI and ML systems? &  IEEE TKDE \cite{wang2020survey}\\
\hline
Edge AI &   What strategies should be employed to oversee the simulation and information transmission among peripheral devices and other systems? What network infrastructures should be utilized to enable this communication?  &  Elsevier IoTCPS \cite{singh2023edge} \& ACM SIGCOMM \cite{ding2022roadmap}\\
\hline
Bitcoin Currency &    How can computing be utilized to maximize the efficiency of computation or processing capacity usage in cryptocurrency for cloud mining? & Elsevier JNCA \cite{ferdous2021survey}\\
\hline
Industry 4.0 &   How can AI, the cloud, and edge computing be used to do predictive analysis that involves company resources? &  IEEE COMST \cite{aceto2019survey}\\
\hline
Intelligent Edge  &  How to deal with big problems that come up when designing system-level, algorithm-level, or architectural-level developments or innovations for integrated cognitive ability, like making decisions in real-time, keeping AI training and inference environmentally friendly, and deploying protection? & IEEE COMST \cite{wang2020convergence} \\
\hline
XAI  &    How can the forecasting of resource and power consumption and SLA variances, as well as the implementation of promptly proactive action, reduce SLA violations and enhance QoS using XAI? &  ACM CSUR \cite{dwivedi2023explainable}\\
\hline
Exascale Computing & How to make energy-efficient computing as power-hungry as the supercomputers that do calculations and transfer data within the computing environment nowadays?  &  ACM CSUR \cite{heldens2020landscape}\\
\hline
6G and Beyond   &  What role 6G may play in reducing latency and improving reaction times by transmitting data between edge devices at high speeds? &  IEEE COMST \cite{shi2023machine} \\
\hline
Quantum AI &    What steps should be taken to build the cloud-based quantum computing infrastructures that are expected to be the foundation for our usage of quantum computers and simulators, which will supplement our existing classical computing hardware? &  Wiley SPE \cite{gill2022quantum} \\
\hline
Quantum Internet & How can the benefits of quantum networking be preserved while integrating the quantum Internet into currently operating conventional technology that will have to exist alongside and communicate effortlessly with today's internet services?  &   IEEE COMST \cite{singh2021quantum}\\
\hline
Analog Computing &   How is it that analog computers can do complicated computations faster and more accurately than their digital equivalents, which utilize ML methods? & Nature Electronics \cite{zhang2020neuro}\\
\hline
Neuromorphic Computing &  How might neuromorphic systems, which model the brain's structure and function and use analog circuits to do AI tasks, pave the way for creating incredibly adaptable, self-learning machines?  & 
 Nature Computational Science \cite{schuman2022opportunities} \\
\hline
Biologically-inspired Computing &  What can researchers take away from brain cells concerning ways to minimize the energy needed for computation, AI, and ML, given that these cells can easily combine smaller tasks to execute larger ones? &  Elsevier ESA \cite{kar2016bio}\\
\hline
Digital Twins &  How can network digital twins aid in speeding up preliminary installations by preparing navigation, protection, digitization, and evaluation in simulation while offering the scalability and interoperability of complex networks? &  IEEE COMST \cite{mihai2022digital} \\
\hline
Net Zero Computing &  How can companies mitigate the negative ecological impact of their IT infrastructure by constructing environmentally friendly data centers and improving energy effectiveness, given that these centers use significant quantities of electricity and release enormous quantities of waste heat while also providing powerful computing services? &   IEEE COMST \cite{cao2022toward}\\
\hline
\end{tabular}
\end{table*}

\section{Impact and Performance Criteria}
 \label{Impact}
In this section, we discuss the impact of contemporary computing and performance criteria. 
\subsection{Performance Metrics} 
We are considering QoS, SLA, autoscaling, and fault tolerance as performance metrics for computing systems.

\subsubsection{QoS and SLA}

Predicting how a cloud computing system will work in real-time is a major difficulty, even if AI techniques are used~\cite{morichetta2023demystifying}. The efficiency of a computer may be measured using QoS metrics, including execution time, cost, scalability, elasticity, latency, and dependability. A SLA, a legally binding contract between a cloud service consumer and provider, defines QoS standards and potential penalties should they be violated~\cite{wright2019performance}. Today, various IoT applications can use blockchain and similar technologies. Each one has its own QoS factors that depend on its area, goal, and demand~\cite{materwala2023qos}. An SLA may also be assessed with a metric called SLA violation rate, which determines compensation in the event of an SLA breach by estimating the divergence of the real SLA compared to the needed (estimated or predicted) SLA~\cite{sharma2023sla}. Since compromized QoS in one cloud service may negatively impact the QoS of the entire computing system, QoS is becoming increasingly crucial while assembling cloud services. Provisioning the proper quantity and quality of cloud resources that will satisfy the QoS of an application's price range, response time, and deadline is essential for providing an effective cloud service~\cite{khan2021guaranteeing}. Consequently, cloud providers should guarantee to offer sufficient resources to minimize or reduce the SLA violation rate, allowing users' workloads to be executed in accordance with their set time and cost constraints~\cite{dilek2022qos}. In that regard, the diversity of applications and their behaviors on different machines requires a tighter description of their needs to minimize SLA violation while not over-provisioning infrastructure~\cite{pujol2023towards}. QoS-aware resource management methods, which can determine and meet the QoS needs of a computing system, such as SLO-driven modeling and execution-reordering of web requests, are crucial to its success in the future~\cite{patros2017slo}. Several research issues must be overcome before QoS can be attained effectively~\cite{singh2017journey}. Initially, the execution time of an application is large, and its performance is diminished due to a lack of cloud resources during runtime---which can be compounded by transparent processes to the developer, such as garbage collection, magnifying the potential of inexplicable SLO violations~\cite{patros2017investigating}.
Additionally, finding the requirement for effective SLA-aware resource management methods decreases the SLA violation rate and preserves the overall efficiency of the computing system. Finally, to reach the ultimate goal of having multiple clouds, there has to be a unified SLA standard across all cloud providers~\cite{zeng2020sla}. Since many IoT applications rely on cloud computing systems that employ AI-based supervised or unsupervised algorithms for learning or models for forecasting, it is imperative to determine the appropriate balance amongst various QoS needs.

\subsubsection{Autoscaling}
Thanks to the dynamic nature of the cloud, self-adapting techniques may be used to reduce resource costs without compromizing QoS~\cite{qu2018auto}. Resource autoscaling, or strategy, reconfiguration, and provisioning, allows for self-additivity. Scientists have looked into autoscaling, or the dynamic modification of computational resources like VMs, for several reasons~\cite{xing2012virtualization}. These include the desire to learn more about (a) horizontal changes, or the addition or removal of VMs; (b) vertical transformations, or the addition or removal of VM resources; (c) choice-making techniques, such as analytical modelling, control theory, and neural networks; and (d) utilizing a range of pricing models, such as on-demand. When it comes to latency-sensitive QoS requirements, the primary challenge for autoscaling methods is figuring out how to make a scaling decision quickly enough. AI prediction is the initial step towards making decisions in the quickest way possible~\cite{du2023computation}. However, traditional ML may not be up to the task when it comes to IoT applications requiring real-time mistake correction due to a lack of autonomous error correction~\cite{lorido2014review}. Also, the rise of latency-sensitive IoT apps and microservices that need responses in the range of milliseconds has made things worse while container-based solutions and burstable efficiency resources should make it possible to deploy and provision resources in the cloud quickly. To prevent a potentially disastrous situation, a smart car's onboard computer constantly monitors data such as the vehicle's speed, the location of other drivers and passengers, and the road conditions~\cite{singh2021rhas}. The cloud alone cannot answer this problem due to the instability and latency in connections between the cloud and users; instead, autoscaling techniques for IoT applications must take these factors into account~\cite{heinze2014auto}. The truth is that autoscaling needs to be made bigger because the cloud naturally gets in the way of Industry 4.0 ideas, like real-time management, and making decisions without a central authority.
\subsubsection{Fault Tolerance}
Providers of cloud computing services owe it to their customers to make such services available without interruption, regardless of what problems arise~\cite{gill2019holistic}. To meet the QoS standards of a computing system efficiently, fault tolerance approaches are employed. Software, hardware, and even networks may all go wrong when a computer system operates. In addition, fault resilience guarantees the reliability and accessibility of cloud services~\cite{gill2022ai}. Timeout breakdowns, overload issues, and resource-lack failures are further examples of cloud dependability issues. A major breakdown has the potential to cause a cascade of failures in the system~\cite{bharany2022energy}. Several proactive and reactive fault tolerance approaches have been developed to cope with these kinds of failures. The most common method of handling faults in long-running processes is called ``checkpointing,'' and it involves preserving the current state after each modification~\cite{gill2018failure}.
Additionally, checkpoints are employed if there is a possibility of not beginning at the same position~\cite{buyya2018manifesto}. Replication-based resilience is another well-known method; it involves duplicating the nodes or jobs until they are completed. If a system is overloaded or malfunctioning, a task migration-based resilience solution can move the work to another computer. Computer systems must have autonomous resilience-aware resource management technology, reliability of service methods, and reliable information integrity (e.g., blockchain) to keep running. Reliability impacts QoS in cloud computing while still delivering it effectively. One of the biggest obstacles in cloud computing is figuring out how to deliver a secure and effective cloud service while cutting down on power consumption and emissions~\cite{gill2020tails}. Cloud computing has built-in redundancy to maintain service availability, QoS, and performance guarantees. Resource management must consider varying failures and workload prototypes for medical care, urban planning, and agricultural applications to run well~\cite{gill2017iot}. Predicting failure in systems that use cloud computing is difficult and can impact the dependability of the system~\cite{gill2018failure}. Predicting faults and achieving the requisite dependability of the cloud service while maintaining QoS necessitates several machine or deep learning approaches~\cite{gill2019transformative}. Replication-based fault tolerance solutions are effective for IoT applications because they reduce task delay and response time. A dependable cloud storage system that will offer an effective retrieval system for processing big data is also required to deal with big data applications~\cite{gill2021manifesto}.
\subsection{Efficiency Metrics}
We are considering energy consumption, carbon footprint, and serviceability as efficiency metrics for computing systems.
\subsubsection{Energy Consumption} 
Data collection and processing have risen exponentially during the last several years. This pattern has been pushing cloud systems to the limits of their computational and, by extension, energy consumption capacities~\cite{katal2023energy}. Annually, CDCs have increased their power use by around 20\% to 25\%~\cite{masanet2020recalibrating}. This shift has led to the rise of decentralized computer architectures such as Fog and Edge. The latency and cost-effectiveness of cloud computing are all vastly improved by moving parts of its computation to distributed edge devices and networks. There nevertheless exist difficulties associated with this. Irregular energy supply, even without the power supply itself, presents significant issues for numerous highly critical and remote sensing applications. 

The ever-growing number of IoT devices and the data they produce have put networking's ability to handle information, compute, and transfer data throughput to the test~\cite{gill2018taxonomy}. Meanwhile, smaller IoT devices are currently created with limited computing power, storage spaces, and energy. Hence, it is imperative to boost the performance of fog and edge nodes in the network. Sustainability in CDCs and minimizing their carbon impact have also become more pressing concerns. This must be accomplished without lowering the bar for QoS~\cite{iftikhar2023hunterplus}. Notwithstanding the obstacles, there have been several advances in this area. Software, hardware, and transitional approaches have all been taken to the energy management problem. 

Approaches and techniques are being designed to optimize software efficiency, supported by computational models~\cite{iftikhar2023hunterplus}. One example is mobile edge computing offloading. Hardware-wise, particularly for the application, devices were designed to provide peak performance while minimizing energy consumption. Energy efficiency in Wireless Sensor Networks (WSNs) has been extensively researched~\cite{gill2022ai}. Fog/edge-node sleep time scheduling, active resource management, and additional energy-saving strategies have all been used in the intermediate phase. There are still many unanswered questions and potential avenues for development when it comes to the effectiveness and longevity of fog, edge, and cloud infrastructures. 

Advanced algorithms for encoding data into fewer bits are explored to reduce transmitter power needs, which are crucial due to limited transmission bandwidth, more critical than direct CPU power needs. Despite the need for specialized hardware, encoding methods may be used by taking advantage of the universal encoders present in virtually all mobile devices~\cite{gill2019transformative}. Yet, it has become impossible to lower the ideal bandwidth due to the rising quantity of data exchange and loss. Preparing for CPU and data utilization in a way that minimizes heat generation requires modelling at the transistor level, which necessitates the development of 3D thermal simulation systems~\cite{tuli2020ithermofog}. 

Lastly, the aim is to minimize power consumption to the point that the CPU and transceiver may be powered entirely by energy harvesting or scavenging approaches~\cite{tuli2022hunter}. Consequently, the Fog/Edge network's granularity may be decreased, leading to more widely scattered, overbearing, and resilient architectures. In various fields, like energy limits, blockchain algorithms might be studied with various versatile AI-based learning approaches for enhanced energy scheduling.
\subsubsection{Carbon Footprint}
End-user needs for applications and the resulting growth in storage in the Exabyte range will result in the first Exascale system by 2025, followed by a Zettascale system by 2035~\cite{lindsay2021evolution}. While this is certainly something to be proud of, there are also many difficulties that come along with it. Keeping everything running requires massive amounts of energy, which poses a major obstacle. At the moment, over ten percent of the world's power is used each year by the ICT sector~\cite{cao2022toward}. The rebound effect, which leads to even higher demand and consumption, makes it counterproductive to create ever-larger systems by increasing efficiency. The next generation of autonomous system paradigms will likely place a greater emphasis on power and carbon footprints in light of climate change and the projected 1.5°C rise in worldwide temperatures owing to emissions of carbon dioxide by 2100~\cite{lindsay2021evolution}. This is not merely about lowering energy use per unit of processing, as is the case now, but also about more basic issues with systems that assume continuous stable power supplies, connectivity with sources of clean energy, and alternate techniques of minimizing energy usage~\cite{schneider2023harnessing}. The study and treatment of systems as living ecosystems rather than as collections of discrete components is a topic of great interest, and this includes the comprehensive integration of managing energy (asynchronous computation, power scaling, wake-on-LAN, air conditioning, etc.).

\subsubsection{Serviceability/Usability}
The fields of human-computer interaction and networked systems have yet to fully merge with each other. This closer synchronization would be especially helpful for cloud computing~\cite{buyya2018manifesto}. Despite significant work on resource management and the back-end associated concerns, accessibility is a vital component in lowering the costs of organizations investigating cloud services and infrastructure. Costs associated with labor might decrease since customers will receive superior service and increase their output~\cite{hartmann2022edge}. 

NIST's Cloud Usability model addresses five dimensions of cloud usability: capability, personalization, reliability, security, and value, all of which have been highlighted as critical issues~\cite{baek2019enhancing}. The term ``capable'' refers to the degree to which cloud service can fulfil the needs of its customers. With the assistance of personal customization options, individuals and businesses will have the capability to modify the visual style and adjust or eliminate features from interfaces for various services. Trustworthy, robust, and useful are attributes associated with possessing a system that fulfils its duties throughout state situations, is safely protected, and delivers value to customers accordingly. Current cloud initiatives have mostly concentrated on wrapping up sophisticated services into APIs that can be accessed by end users~\cite{hartmann2022edge}. HPC Cloud is the most evident example. To make HPC applications more accessible and easier to use, researchers have developed several different services. In addition to being packaged as services, these systems provide Web interfaces through which their settings may be set and their input and output files managed. 

DevOps is another path associated with cloud usage that has gained popularity in recent years~\cite{miraz2021adaptive}. DevOps has increased the efficiency of both software engineers and administrators when it comes to developing and delivering remedies on the cloud. Cloud computing is important not only for creating brand new solutions AIOps and MLOps~\cite{diaz2023joint} but also, for streamlining the process of moving existing applications from onsite settings to adaptable, multi-tenant cloud services.

\subsection{Social Impact}
We are considering the digital divide, ethical AI, and digital humanism as social impact metrics for computing systems.

\subsubsection{Digital Divide}
Corporations in rural areas have significant challenges due to the difficulty of gaining a connection to broadband connectivity and, by extension, cloud-based resources~\cite{celik2023exploring}. Access to the web is one example of a long-standing infrastructural gap between urban and rural areas. There are a lot of companies that can not expand and innovate because they lack access to new technology. Businesses in rural areas face another obstacle: the high cost of maintaining and upgrading on-premises IT infrastructure. Cloud computing's main benefits are the ability to work together and think creatively. The cloud encourages teamwork by facilitating real-time, distributed collaboration. This greater collaboration encourages invention. As a result, rural enterprises may now compete on an equal basis with their metropolitan competitors~\cite{gill2024transformative}. Accessibility to data and fundamental information is also crucial. The benefit of using the cloud has increased significantly with the advent of generative AI. Comprehensive sales, marketing, and manufacturing capabilities are provided by core AI services, but these cannot be reproduced with human processing and can be too costly to install on-site for modest organizations. The proliferation of cloud computing has expanded business opportunities, but not equally. By utilizing the cloud, companies in rural areas may overcome the constraints of their physical location~\cite{le2011can}. Cloud computing's greater availability, decreased cost, scalable effectiveness, and improved cooperation may breathe new life into the rural economy and propel it towards long-term success.

\subsubsection{Ethical AI}
AI systems require vast amounts of data, including details on businesses and their clients~\cite{arce2024optimizing}. The value of knowing the data owner surpasses that of having private information that cannot be linked to a specific person. When dealing with sensitive information, companies regularly face problems related to data security and regulatory compliance~\cite{qadir2022toward}. Autonomic computing using AI needs to take into account privacy rules and data protection. While AI has the potential to be a game-changer, it has not always been successful in achieving its aims. A hunt for answers by an AI may result in a flood of insensitive comments~\cite{munn2023uselessness}. The vast number of AI decisions and the stakes involved make this field fraught with peril. Prior to expanding the use of this invention, it is crucial to develop accountability and ownership.
\subsubsection{Digital Humanism}
The unavoidable consequences of digital colonization driven by business need a counter-force of digital humanism motivated by care for humanity and the Earth~\cite{scuotto2023digital}. We have never been both so interdependent, yet so isolated. Modern digital systems allow for global communication. One no longer has to be in the same room as someone else to have a conversation, collaborate on a project, or just have fun with them. The cell phone is rapidly becoming an integral part of people's daily lives all across the world. Connectivity between the developing world and the developed nations of the world is rapidly expanding, for both good and ill. These interconnections are causing conflicts that could have been prevented when individuals and ideas were separated by space. Western materialism and commerce meet Eastern spirituality and culture in the virtual world~\cite{schaap2017gods}. Therefore, although humans may all end up in the cloud at some point, the barriers of mutual respect and compassion that keep us from crashing into one another are more than frayed. Most modern digital accounting and tracking systems are used by private companies seeking to maximize profits at the expense of others, enriching a few elites at the expense of a much larger underclass~\cite{magni2023digital}. In contrast, if the cloud could be utilized for humanity's benefit, manufacturing and distribution might be dramatically enhanced. Controlled well, such instruments will allow for fine-tuning of many crucial societal functions, particularly at the subnational and neighborhood levels.
\subsection{Security and Compliance}
We are considering data protection, privacy regulations, and resilience to attacks as security and compliance metrics for computing systems.
\subsubsection{Security, Privacy and Resiliency}

In recent years, there has been a dramatic change in academia and business towards the IoT, edge computing, and cloud computing in order to serve customers better. With this massive paradigm shift, comes a slew of problems and difficulties with protecting the confidentiality and safety of the information stored on these devices~\cite{yu2019lagrange}. Edge computing's many distinguishing features---its low latency, geographical dispersion, end-device accessibility, high processing power, variability, etc.---make it imperative that security and privacy mechanisms be both flexible and powerful~\cite{olowononi2020resilient}. In addition, creating universally compatible software platforms is challenging due to the wide variety of use cases and device types. 

Several elements become important in the research of these security and associated challenges in the cloud and fog computing models: End-user confidence and privacy; verification and validation of sources inside nodes; secure communications between sensor, compute, and broker nodes; detection and prevention of malicious attacks; secure, reliable and decentralized data storage, such as Blockchain~\cite{golec2021ifaasbus}. Some of the problems that have already been addressed in this field include adaptive mutual authentication, identifying and retrieval of harmful or malfunctioning nodes, the detection and defence against assaults, the avoidance of harmful hazards, and the protection of user information from theft. Unmanned Aerial Vehicle (UAV)-aided computing devices can now maintain their anonymity while contributing to distributed frameworks in AI technology, such as computer vision and path learning, supporting data processing and decision-making~\cite{liu2022efficient}. Other efforts in fog forensics have also given digital evidence by recreating prior computer activities and identifying how these events contrast with cloud forensics in important ways. 

The past few years have seen significant progress in several key areas related to Fog Radio Access Networks (F-RANs), including mobility management, interference reduction, and resource optimization~\cite{samriya2023secured}. Novel approaches have evolved for varied applications handling privacy challenges. Face recognition and resolution, vehicle crowd sensing, geographic location sensing and data processing, renewable node storage systems and data centers, and fog-based public cloud computing are promising new research areas. Prevention against data theft, attacks involving man-in-the-middle, confidentiality of users, location confidentiality, forward privacy, reliable user-level key management, and many other weaknesses have all been addressed through such efforts~\cite{gill2022ai}. 

There are scaling issues with many fog/cloud privacy and security models that prevent them from fully applying to the next-generation edge computing transition~\cite{ullah2023privacy}. Because of fog computing's decentralized nature, numerous new security concerns, which are not an issue in the cloud, emerge in the fog layer and IoT devices. The deployment of authentication systems is hampered by the prevalence of threats such as advanced persistent threats (APT attacks), malware, distributed denial of service (DDoS) attacks, two-way communication, and micro-servers without hardware protection mechanisms in edge data centers~\cite{kim2020resilient}. Additionally, these studies show how the mobile edge computing architecture might change in the future. For example, edge nodes working together could make real-time encryption more efficient. The computational capacity of both edge and distant resources has not been completely used in previous efforts, and security flaws have been addressed from a restricted viewpoint. New phenomena appear when cloud-like capacities are distributed to the network's periphery~\cite{golec2021ifaasbus}. Edge data center collaboration, service migration on a local and global scale, end-user concurrency, QoS, real-time applications, load distribution, server overflow issues, stolen device detection, and dependable node interaction are all examples of such scenarios. Future studies can focus on new areas, such as evolving game-theoretical strategies to the privacy algorithms encouraged by adversarial attack scenarios, communication protocols in sensor cloud systems, and clustering model-based security evaluation (AI-based forecasting approaches), which can be investigated as potential solutions to these issues~\cite{weichbroth2023security}. Mobile devices' presence in these data centers should be taken into account by safeguarding systems.

\subsection{Economic and Management}
We are considering cost-efficiency, resource allocation, application design, computing economics, and data management under economics and management for computing systems.
\subsubsection{Cost-Efficiency}
Minimizing cloud expenditures while maximizing application performance and efficacy is the goal of cloud cost optimization, which entails striking a fine balance between technological standards and corporate goals~\cite{katal2023energy}. Cost-effective cloud computing refers to the practice of utilizing cloud providers in the most economical way feasible to operate software, complete tasks, and create value for a company. Optimization as a practice varies from fundamental business management to challenging scientific and technical fields including operational research, statistical and data analysis, and modelling and prediction~\cite{arce2024optimizing}. Corporations may maximize the return on their investments in cloud computing through cost optimization, which reduces wasteful expenditures and strengthens their operational effectiveness~\cite{delacour2023energy}. By avoiding economic hazards, aligning spending with company goals, and establishing a secure, scalable, and cost-effective cloud infrastructure, corporations can maximize the return on their investments in cloud computing. In general, efficient cloud cost management preserves essential resources against the risk of unanticipated expenditures and financial mismanagement. Changing to a cloud-native methodology involves more than just updating technology; it also necessitates a substantial adjustment in mindset~\cite{buyya2018manifesto}. Building scalable apps that make efficient use of resources requires developers to think in terms of the cloud from the start. To optimize cloud expenditures, a cloud-native application design requires an in-depth familiarity with the services and resources offered by different cloud service providers. Managed service options are superior to autonomous technologies since they require less effort and time investment~\cite{quan2025historical}. A sophisticated knowledge of the user application's demands, regulatory demands, and possible financial consequences is necessary to choose between a single and multi-cloud installation plan. An organization's administration might be simplified by adopting a single-cloud approach, but doing so could leave it vulnerable to vendor lock-in and service restrictions~\cite{lindsay2021evolution}. Contrarily, a multi-cloud strategy can increase complexity in administration but has the ability to optimize costs, provide greater flexibility, and lessen the danger of vendor lock-in. Identifying which is the most economical and profitable implementation approach requires careful consideration of the specific features, pricing methods, and competencies of different cloud services.
\subsubsection{Resource Allocation}
The sheer size of today's CDCs makes resource management in networked systems a formidable challenge. In large-scale distributed architectures, the variety of network devices, elements, and ways to connect raises the difficulty of resource management strategies~\cite{mijuskovic2021resource}. Consequently, there is a necessity for innovative resource allocation methodologies that would add to the reliability and effectiveness of these systems while keeping them cost-effective and sustainable.
While resource management is fundamental to distributed systems (be it the cloud, the IoT, or fog computing), additional guarantees are needed to ensure that these systems operate well in terms of latency, dependability, cost-effectiveness, and throughput~\cite{singh2016survey}. The software layer is just one part of these larger systems, which also require consideration of networking, server architecture, and ventilation. By incorporating blockchain technology into operations like resource sharing and VM migration, cloud systems may be more secure~\cite{hong2019resource}. There is a pressing need to investigate novel approaches to managing computer system resources by taking a systemic perspective and using AI models. Moreover, experiment-driven strategies for examining methods to optimize resource management methods may be investigated~\cite{jamil2022resource}. Borg was opened up by Google as Kubernetes, which is an instance of a cluster management system that incorporates data abstraction into resource management. Users are freed from worrying about the nuts and bolts of resource management and may instead focus on composing cloud-native applications. 

Borg conceptually separates the whole cluster into cells, each housing a Borgmaster (controller) and a Borglet (which initiates and terminates tasks within the cell's perimeter). The master node coordinates with the Borglets and processes RPCs from clients requesting actions like creating jobs or reading data~\cite{aslanpour2020performance}. This centralized design is very suitable for scaling. The primary benefit of this architecture is that operations that have already been started will continue to execute even if the master or a Borglet fails~\cite{raju2019comparative}. 

A system known as Mesos can facilitate the equitable distribution of commodity clusters. It coordinates the use of commodity clusters by many systems. The fundamental idea is to make use of available resources~\cite{henning2024benchmarking}. In this model, Mesos determines how many resources to give to every framework depending on the limitations associated with that framework, and the frameworks then choose which offers to take. Thus, scheduling choices must be made by frameworks. In addition, Mesos facilitates the creation of domain-specific frameworks (like Spark) that may greatly enhance performance. To schedule and manage available resources, YARN is used as a framework~\cite{buyya2018manifesto}. It enables services to ask for computing power at various topological levels, including individual servers, networks, and whole racks. The primary component in charge of allocation is YARN's resource management. Similarly to Mesos, it enables several frameworks to collaborate on the same commodity clusters~\cite{raju2019comparative}. YARN's integrated reliability masks the complexities of failure identification and recovery.

\begin{itemize}

\item \textbf{Heterogeneous Resources and Workloads: } There is a lack of cohesion in the existing literature about managing resources and workloads in diverse cloud settings. As a result, there is no common setting in which cloud applications can make optimal use of heterogeneity in VMs, vendors, and hardware architectures~\cite{rosenfeld2022query}. Consequently, the initiative recommends an overarching program that takes into consideration diversity throughout. Effective solutions can be picked from a collection of workload and resource handling methods, depending on an application's needs~\cite{feng2022heterogeneous}. Heterogeneous memory control is necessary for this purpose. Modern memory control techniques rely heavily on hypervisors, thereby minimizing the potential advantages of heterogeneity. Recent calls for action have advocated for alternatives that focus on heterogeneity awareness in the guest OS. Another chasm is that between heterogeneity and abstraction~\cite{garofalo2022heterogeneous}. Accelerator-specific languages and low-level programming initiatives are necessary for today's programming paradigms to utilize hardware processors. Furthermore, such models allow for the creation of useful research software. As a result, service-oriented and user-driven applications on cloud platforms are hampered in their ability to take advantage of heterogeneity. Kick-starting an international community initiative to come up with an open-source, high-level programming language that is suitable for cutting-edge and creative Web-based applications in a heterogeneous setting is a worthwhile step to take~\cite{wu2020collaborate}. Whenever fog computing matures and application migration occurs, such aids will be invaluable.

\end{itemize}

\subsubsection{Application Design}
By 2025, analysts predict 61 billion connected devices will generate 40 percent of global data at the cost of \$2.5 trillion~\cite{kumar2023digital}. Medical services, near-real-time traffic management systems, precise farming, intelligent towns and cities, etc., are just a few examples of IoT applications that are driving the need for improved processing capacity, data storage, confidentiality, security, and trustworthy communication. Additionally, as the data produced by these devices is used to resolve real-time challenges, credibility, uniformity, and accessibility of the data must be maintained. It's challenging to design such complex applications for IoT systems~\cite{sha2020survey}. As a result, it is essential to develop application designs and architectures that are not only dependable and quick enough to deliver effective efficiency but additionally, scalable to manage massive amounts of data through these devices. These are the most important factors to consider when developing such apps for cloud environments. Firstly, a data packet's latency is the time it takes to travel between an IoT device and the cloud before returning. For time-sensitive information, even a millisecond delay might have drastic consequences. For instance, having a crisis-sensing instrument that only sounds an alert after a disaster has already taken place is not a viable solution. Data needing immediate reaction should be analyzed as close as possible to the origin~\cite{sequeiros2020attack}. Secondly, if all this data is transferred to the cloud for storage and analysis, the resulting traffic will be massive, using up all available bandwidth. The distance between the device and the cloud also increases transmission latency, which slows down responses and reduces user experience. Therefore, some tasks must be transferred from the cloud to an edge server located between the Internet servers and the mobile device: such solutions better satisfy end-users' requirements.

By storing and processing certain IoT data directly on IoT devices, the fog computing model reduces the load on the cloud and keeps costs down. Large-scale, geographically dispersed applications that rely heavily on real-time data benefit from the fog's consistency~\cite{kaur2018future}. Fog computing may be the most appropriate choice to enable effective IoT and provide reliable and safe services and resources to many IoT users. Big data analytics, IoT devices, fog, and edge computing have become the foundations for smart city programs worldwide~\cite{sebastian2020memory}. In transport, fog computing is useful for several tasks, including vehicle-to-vehicle interaction, smart-sensor-based congestion control system management, driverless car management, and self-parking, among others. Furthermore, governments may employ these applications to make the lives of their residents safer and more environmentally friendly, making them a sustainable approach. Emergency services, such as those dealing with fires or natural disasters, can also benefit from this technology by receiving timely alerts about developing crises to help them make informed choices.

Farming software that tracks weather and climatic data like rainfall, wind speed, and temperatures, makes it easier for farmers to reap a harvest. An IoT agriculture platform is suggested for cloud and fog computing, with applications including automated agricultural monitoring, visual inspection for pest control, and more efficient use of farm resources~\cite{sha2020survey}. Meanwhile, in the medical field, more and more people are using fitness trackers, blood pressure monitors, and heart rate monitors to track vital signs and gather data for medical analysis. Thanks to these innovations, physicians can check their patients' health from afar, and patients have more say in their care and decisions.

\subsubsection{Computing Economics}
There are several promising new avenues for study in the financial aspects of cloud computing. It is becoming clearer that the lower costs of container deployment can be used to handle real-time workloads~\cite{vu2020ict}. This is speeding up the switch from VMs to containers for cloud computing. 

\begin{itemize}

\item\textbf{Cost-Effective Computing Models:} In serverless computing, no billing for computing resources is made until a function is invoked. Processes executed in these lambda functions tend to be narrower in focus and designed for processing data streams. Whether or not serverless computing is beneficial for a given application depends on its projected runtime behavior and workload~\cite{buyya2018manifesto}. Averaged versus peak transaction rates; scaling the number of simultaneous operations on the infrastructure (i.e., operating multiplies simultaneous functions with a growing number of consumers); and benchmark implementation of serverless functions across various backend hardware platforms~\cite{tesfatsion2023agent}. Conversely, increased employment of fog and edge computing characteristics with cloud-based data centers gives tremendous study potential in cloud economics.

\item \textbf{Economic Impact of Computing Technologies: }It is possible to lower the expenses of running cloud services and infrastructure by combining reliable resources of the cloud with more ephemeral resources at the consumer's edge. To make such technology accessible at the edge, nevertheless, it is anticipated that consumers will require some sort of inducement~\cite{buyya2009cloud}. Expanding the cloud market to include new types of service providers is possible because of the accessibility of cloud and edge resources. Researchers call these intermediate facilities located between the conventional data center and the user-owned or provisioned resources, microdata centers~\cite{vairetti2024analytics}. The federation concept in computing allows for many microdata center operators to operate together to distribute workloads in a given region at desired pricing.
\end{itemize}

\subsubsection{Data Management}

Metadata handling for datasets is not given much attention in cloud IaaS and PaaS services for storing and information administration, which instead prioritize file, partially structured, and structured data separately. In contrast to traditional, organized data warehouses, proponents of ``Data Lakes'' advocate for businesses to store all their data in unstructured formats on the cloud, using services like Hadoop~\cite{buyya2018manifesto}. Nevertheless, using them might be difficult due to the absence of information for tracking and defining the origin and authenticity of the data. 

\begin{table*}[!t]
\caption{Summary of open challenges and future directions in the above-discussed impact of modern computing and performance criteria with future reading}
\label{tab:table4}
\centering
%\footnotesize
\begin{tabular}{p{2.6cm}|p{10.5cm}|p{3.0cm}}
\bottomrule
\textbf{Impact and Performance Criteria} & \textbf{Open Challenges and Future Directions} & \textbf{Further Reading}\\
\hline
QoS and SLA &  How can SLAs and QoS be preserved in real-time when cloud computing and edge resources and tasks are executed? & ACM CSUR \cite{sharma2023sla} and Wiley IJCS \cite{dilek2022qos}\\
\hline
Autoscaling  &  How can it be ensured that computing resources need to meet SLAs and QoS are effectively autoscaled in real-time?  & ACM CSUR \cite{qu2018auto}\\
\hline
Fault Tolerance &    How can reliable support be continuously provided with environmentally-friendly services? & Elsevier SETA \cite{bharany2022energy}\\
\hline

Energy Consumption   &  How can modern computing benefit from AI/ML to provide environmentally-friendly services and low energy consumption?  & Springer Cluster Computing \cite{katal2023energy}\\
\hline

Carbon Footprint & What technological advancements may decrease the impact of climate change and how could environmentally-friendly computing have a lower-carbon footprint? &  IEEE COMST \cite{cao2022toward} \\
\hline

Serviceability  &   What methodologies should be employed to develop and measure key performance indicators, also known as KPIs, in order to assess the success of initiatives that aim to make cloud computing more usable and secure? & Wiley ETT \cite{hartmann2022edge}\\
\hline

Digital Divide & How does the use of the cloud help overcome the digital divide? Can ICTs help bridge the digital divide in infrastructural growth? &   Elsevier Telematics and Informatics \cite{celik2023exploring} \\
\hline

Ethical AI &  When designing and implementing AI in computing devices, what ethical concerns must be taken into account? & Nature Machine Intelligence \cite{jobin2019global} \\
%Emerald JICES \cite{qadir2022toward}\\ %Springer AI and Ethics \cite{munn2023uselessness} \\
\hline

Digital Humanism & How may digital tools stimulate original thought and the independent thinking of individuals, and whether or not the synergy of these traits can promote a digital shift in the workplace?  & Elsevier Journal of Business Research \cite{scuotto2023digital} \\
\hline

Security, Privacy \& Resiliency  &  What measures can be taken to ensure that personal information is protected and data is securely processed in the cloud when IoT apps collect and analyze massive amounts of data?  & IEEE COMST \cite{olowononi2020resilient} \\
\hline

Cost-Efficiency & How can impending difficulties like the prohibitive cost of setting up and running big systems testing environments and the influence of global warming on the architecture of upcoming systems be overcome? &  Springer Cluster Computing \cite{katal2023energy}\\
\hline

Resource Allocation  &  What are the best practices for successfully provisioning cloud and edge resources for many IoT apps before scheduling such resources?  &  ACM CSUR \cite{jamil2022resource}\\
\hline

Heterogeneous Workloads/ Resources  &    How can the heterogeneity of resources and workloads impact the efficiency of a computing system at runtime? & ACM CSUR \cite{rosenfeld2022query}\\
\hline

Application Design &  How can more efficient IoT apps be developed to make greater use of available computer power? &  ACM CSUR \cite{mahmud2020application}\\
\hline

Computing Economics  &   How can businesses strengthen their CapEx (Capital Expenditure) and OpEx (Operational Expenditure) strategies by learning about the primary economic advantages of cloud computing in terms of return on investment (ROI), total cost of ownership (TCO), and relocation? & Elsevier Telecommunications Policy \cite{vu2020ict}\\
\hline
Data Management  &  How can organizations make optimal use of AI/ML approaches for enormous amounts of data to ensure efficient data administration and analysis?  & Springer JBD \cite{hariri2019uncertainty} \& ACM CSUR \cite{cao2017data}\\
\hline

\end{tabular}
\end{table*}

\begin{figure*}[t]
	\centering
	\includegraphics[scale=0.75,trim=0.5in 0.5in -0.3in 0in]{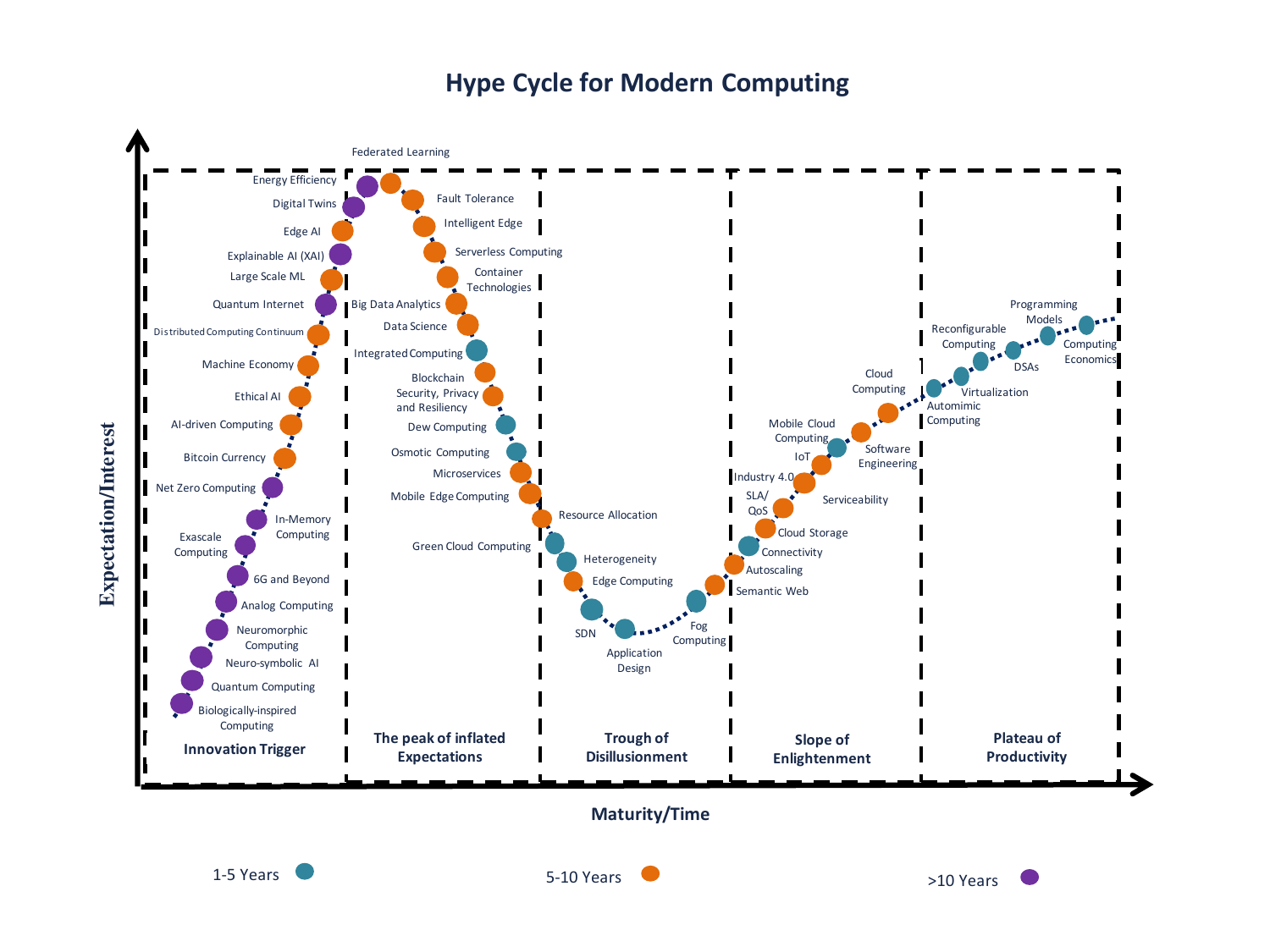}
 \caption{Hype Cycle for Modern Computing}
	\label{fig:hype}
\end{figure*}

Throughout the past ten years, research archives have become exceptional in handling vast, varied datasets and the accompanying information that provides context for their usage. Collocating data and computing resources in a small number of strategically located data centers worldwide allow for economies of scale, a major advantage of CDCs~\cite{hariri2019uncertainty}. Nevertheless, bandwidth restrictions across worldwide networks and delays in gaining access to data present obstacles~\cite{daniel2019big}. This becomes an increasingly pressing issue as IoT and 5G mobile networks expand. However, the cloud providers' access to private data and critical confidential information still poses a risk for businesses that need to guarantee strict privacy for their end-users. Likewise, there are no foolproof auditing techniques to prove that the cloud service provider has not obtained the data, even though regulatory measures are in place. In a hybrid setup, customers may handle confidential information under their watchful eye while still taking advantage of the advantages of public clouds, thanks to the proximity of private data centers to public CDCs connected by an independent high-bandwidth network. Furthermore, effective approaches to managing resource flexibility in such contexts should be explored~\cite{donta2023governance}. In addition, it is preferable to have high-level programming abstractions and bindings to platforms that can allocate and oversee resources in these massively dispersed settings. 

Finally, with the IoT, deep learning, and blockchain all set to be housed on clouds, it's important to look at specialist data management services to ensure their success~\cite{ur2019role}. As indicated above, IoT will include a strengthened requirement to deal with streaming data, their effective storage, and a requirement to integrate data management on the edge effortlessly with administration in the cloud~\cite{gill2019fog}. When unregulated edge devices are involved, integrity and authenticity become even more crucial. As the use of deep learning grows, it will become more important to be able to manage trained models well and make sure they can be quickly loaded and switched between to make online and distributed analytics applications possible~\cite{cao2017data}. Finally, blockchain and decentralized ledgers can improve data management and tracking by providing greater transparency and auditability. While initially used by the financial sector (of which cryptocurrencies are only one prominent example), these systems may be expanded to store other company data safely with an inherent auditing record.

\textbf{Summary:} Table \ref{tab:table4} lists the summary of open challenges and future directions in the above-discussed impact of modern computing and performance criteria, along with recommendations for future reading.

\section{\textcolor{black}{Emerging Trends in Modern Computing}}
\label{Emerging}
\textcolor{black}{The advent of modern computing technology has made it possible to resolve several real-world issues, including delayed responses and low latency. It has facilitated the development of start-ups led by promising young minds from all over the world, providing access to massive computing capacity for tackling difficult issues and accelerating scientific advancement. Thanks to its ground-breaking improvements in efficiency in domains like neural networks, Natural Language Processing (NLP), and related applications, AI has been gaining popularity lately. Computing is a vital infrastructure for running AI services due to its enormous processing power, and AI has the potential to improve existing computing by making resource management effective. Several AI models rely on outside data sets and large-scale computer capacity, both of which might be easier to access with today's computing systems. Currently, training advanced models of AI in large numbers is becoming even more crucial. Additionally, extensive application of AI in contemporary computer systems may be possible due to ground-breaking XAI research. In the decades to come, AI will place substantial stress on computing resources. To meet these demands, it's necessary to develop new approaches to research and methodology that make use of AI models to solve problems with adaptability, delay, and handling of resources and cybersecurity. Scalability and adaptability are two open issues that have not yet made full use of AI models as an economical way to boost the performance of computer applications.}

\textcolor{black}{Our analysis has led us to categorize certain areas of computing into three separate maturity levels: a period of five to ten years, over a decade, and under five years. Several novel innovations are on the horizon that might significantly improve the utilization of modern computing, and the article has highlighted them all over the coming decade. Figure~\ref{fig:hype} depicts the hype cycle for modern computing systems along with their new trends. Researchers extensively study computing paradigms and technologies, with edge AI and federated learning now dominating. New areas of study within computing, such as distributed computing continuum and AI-driven computing are just scratching the surface. Applications for computing in these domains may not mature for another five to 10 years. Quantum ML, sustainability, Net Zero Computing, XAI, and the quantum Internet are all expected to be in the spotlight for at least another decade. Digital twins, cyber security, edge intelligence, edge computing, and blockchain technology have generated an unprecedented level of excitement. They are expected to be completely built-in under five years with the help of modern technology. Machine Economics, In-Memory Computing, Bitcoin Currency and AIOps/MLOps have all reached their peak of inflated expectations for the following five to 10 years of noteworthy evolution. Significant progress needs to be made before biologically inspired computing, neuro-symbolic AI, analog computing, neuromorphic computing, 6G, and quantum computing can be considered hype-worthy. Cloud and fog computing has been trending heavily over the past few years, and that trend could persist for the next five to ten years.}

\section{Summary and Conclusions}
\label{Conclusion}
\textcolor{black}{This research offers a comprehensive exploration of the evolution of modern computing systems over the past sixty years, tracking the transition from classical computers to quantum computing and examining their key components, such as physical architecture, conceptual units, and communication methods. We analyze the influence of conceptualization and physical models on the shift from centralized to decentralized structures, a significant change since the Internet's inception. Developments in microcontroller architecture, operating system design, and networking infrastructure have given rise to ubiquitous computing models like the Internet of Things (IoT), pushing the boundaries of both physical and conceptual realms. The move towards specialized hardware and software, particularly in data-driven fields like AI, represents a shift from earlier focuses on system flexibility and adaptability. This article also addresses issues of accessibility and potential inequalities, emphasizing the need to ensure these technologies positively impact society and everyday life. Integrating recent advancements with ongoing challenges in the application of established technological trends, this work provides an in-depth analysis of the next wave of scientific research in computing. It summarizes current findings, acknowledges limitations, and outlines new trends and key challenges, considering the impact of emerging trends and envisioning future research paths in modern computing. This review aims to be a valuable resource for experts, technologists, and academics interested in the latest developments and future directions in the field of modern computing.}

\section*{Acknowledgements}
\textcolor{black}{We thank the Editor-in-Chief (Prof. Ke Xue) and anonymous reviewers for their insightful comments and recommendations to improve the overall quality and organization of the article. We would also like to express our gratitude to Neil Butler (CEO, CloudScaler, UK), Marco AS Netto (Microsoft Azure HPC, USA) and Manmeet Singh (University of Texas at Austin, USA) for their thoughtful remarks and valuable suggestions.}

\section*{Conflict of Interest}
On behalf of all authors, the corresponding author states that there is no conflict of interest.

\section*{Data availability}
No data was used for the research described in the article.

\section*{\textcolor{black}{Appendix A. List of Acronyms} }
\textcolor{black}{Table \ref{AD} shows the list of acronyms. }

\begin{table}[ht]
\caption{List of Acronyms} \label{AD}
\begin{tabular} {| p{1.6cm} | p {6.2cm}|}   \hline 
\textbf{Abbreviation} & \textbf{Description} \\  \hline
PCs  &   Personal Computers \\
DNS   & Domain Name System  \\
MPP   & Massive Parallel Processing \\
AI   &  Artificial Intelligence \\
SMP    &  Symmetric Multi Processing  \\
OS         &   Operating System\\
GUI      &   Graphical User Interfaces \\
IoT & Internet of Things \\
 HTTP & Hyper Text Transport Protocol   \\
HTML  &  Hyper Text Markup Language \\ 
WWW     &  World Wide Web \\
RPC & Remote Procedure Calls  \\
JSON  &  JavaScript Object Notation \\
XML & Extensible Markup Language  \\
SOA & Service-Oriented Architecture   \\
CDC  &  Cloud Data Centers \\
 HPC & High Performance Computing \\
IT & Information Technology  \\
SaaS &  Software as a Service  \\
PaaS &  Platform as a Service \\
IaaS & Infrastructure as a Service \\
SBC & Single-board Computers  \\
SDN & Software-Defined Networking   \\
NVF &  Network Function Virtualization \\
IIoT  & Industrial Internet of Things   \\
QoS & Quality of Service \\
IoE &  Internet of Energy  \\
B5G &  Beyond 5G \\ 
SLA &   Service-Level Agreement   \\
FPGA &  Field-Programmable Gate Arrays \\
ASICs &  Application-Specific Integrated Circuits  \\
GPU &  Graphics Processing Units \\
CUDA & Compute Unified Device Architecture  \\
TPU & Tensor Processing Units  \\
ICT &   Information and Communication Technology\\
CaaS &   Container as a Service\\
QoE &  Quality of Experience \\
V2X &  Vehicle-to-Everything \\
MEC &  Multi-access Edge Computing \\
VM & Virtual Machines    \\
M2M &  Machine-to-Machine \\
PoW &   Proof of Work \\
XAI &   Explainable Artificial Intelligence\\
UAV & Unmanned Aerial Vehicle  \\
DDoS &   Distributed Denial of Service\\
STCO & Systems-Technology  Co-Optimisation  \\
SoC & System-on-a-Chip  \\
ML  &  Machine Learning  \\
SLO &  Service Level Objective \\

\hline 

\hline 
\end{tabular}
\end{table}

\bibliographystyle{ieeetr}

\bibliography{cas-refs}

\end{document}